\documentclass[letter]{aa} 

\usepackage{graphicx}
\usepackage{txfonts}
\usepackage{hyperref}

\usepackage{float}

\begin{document}

\title{Possible Hycean conditions in the sub-Neptune TOI-270~d}

   \author{M\aa ns Holmberg\inst{1}
          \and
          Nikku Madhusudhan\inst{1}%\fnmsep
          }

   \institute{\inst{1} Institute of Astronomy, University of Cambridge, Madingley Road, Cambridge CB3 0HA, UK\\
              \email{nmadhu@ast.cam.ac.uk}
              }

 \date{}

     \abstract
   {
   The JWST has ushered in a new era in atmospheric characterisations of temperate low-mass exoplanets with recent detections of carbon-bearing molecules in the candidate Hycean world K2-18~b. We investigated JWST observations of the TOI-270 system, with two sub-Neptunes simultaneously transiting the nearby M dwarf during the visit. We report our atmospheric characterisation of the outer planet TOI-270~d, a candidate Hycean world, with JWST transmission spectroscopy using the NIRSpec G395H instrument in the 2.7-5.2 $\mu$m range, combined with previous observations obtained with the HST WFC3 spectrograph (1.1-1.6 $\mu$m). The spectrum reveals strong signatures of CH$_4$ and CO$_2$ at 3.8-4.9$\sigma$ and 2.9- 3.9$\sigma$ confidence, respectively, and no evidence of NH$_3$. The abundant CH$_4$ and CO$_2$, at $\sim$0.1-1\% mixing ratios, and the non-detection of NH$_3$ are similar to the findings reported for K2-18~b and consistent with predictions for a Hycean world with a planet-wide ocean under a H$_2$-rich atmosphere. We also report evidence of CS$_2$ at a 2.3-3.0$\sigma$ confidence and a potential inference of H$_2$O at 1.6-4.4$\sigma$, depending on the data analysis approach, and discuss possible interpretations of these results. The spectrum does not provide strong constraints on the presence of clouds or hazes in the observable atmosphere, nor any evidence for the effects of stellar heterogeneities, which is consistent with previous studies.  For the smaller inner planet TOI-270~b, we find that the spectrum is inconsistent with a featureless spectrum at $\sim$3$\sigma$, showing some preference for an H$_2$-rich atmosphere in a super-Earth. We discuss the implications of our findings and future prospects. 
   }
   

   \maketitle

\section{Introduction} \label{sec:intro}

We have entered the era of precision remote sensing of temperate low-mass exoplanets with JWST. Planets in the sub-Neptune regime (radii of 1-4 $R_\oplus$), with no analogue in the solar system, span a wide range of possible interior and atmospheric compositions -- from predominantly rocky super-Earths to volatile-rich mini-Neptunes. Recent detections of carbon-bearing molecules in the atmosphere of a possible Hycean world \citep{Madhu_carbon2023} have opened the door to similar detections in other temperate sub-Neptunes orbiting nearby M dwarfs. Hycean worlds are a class of temperate sub-Neptunes with planet-wide habitable oceans underlying shallow H$_2$-rich atmospheres \citep{Madhusudhan2021}. In particular, several such planets have been proposed as promising candidates for Hycean conditions and are conducive to atmospheric characterisations with JWST.

The TOI-270 system is widely known as an excellent target for atmospheric characterisation of sub-Neptunes \citep{Chouqar2020, Kaye2022}. TOI-270 is a nearby (22.5 pc), quiet, M3V star that hosts three confirmed transiting sub-Neptune planets \citep{Gunther2019}. The innermost planet, TOI-270~b, has a radius of 1.21$\pm$0.04 $R_\oplus$ and a mass of 1.58$\pm$0.26 $M_\oplus$. The two outer planets, TOI-270 c and d, are larger with radii of 2.36$\pm$0.06 $R_\oplus$ and 2.13$\pm$0.06 $R_\oplus$, respectively, and masses of 6.15$\pm$0.37 $M_\oplus$ and 4.78$\pm$0.43 $M_\oplus$, respectively \citep{VanEylen2021}. Both planets have been identified as candidate Hycean planets, however, TOI-270 d, with a lower equilibrium temperature, is a more promising Hycean candidate and an optimal target for atmospheric characterisation in the temperate regime \citep{Madhusudhan2021}. This planet was recently observed with the Hubble Space Telescope (HST), suggesting an H$_2$-rich atmosphere with evidence of H$_2$O absorption in the transmission spectrum \citep[][]{Evans2023}, making it an interesting target for further observations with JWST.

Spectroscopic observations with JWST have the potential to break the degeneracy between Hycean conditions and other atmospheric scenarios and provide important insights into the atmospheric, surface, and interior conditions of a planet. Several studies have demonstrated that atmospheric characterisation can be used to constrain a surface beneath H$_2$-rich atmospheres in temperate sub-Neptunes \citep{Yu2021, Hu2021, Tsai2021, Madhusudhan2023}, such as the case of K2-18~b \citep{Madhu_carbon2023}. For Hycean worlds, the presence of an ocean below a thin H$_2$-rich atmosphere may be inferred by an enhancement of CO$_2$, H$_2$O, and/or CH$_4$, together with a depletion of NH$_3$ \citep{Hu2021, Tsai2021, Madhusudhan2023}. Therefore, for the Hycean candidate TOI-270~d, observations of these key  carbon-, nitrogen-, and oxygen- (CNO) bearing molecules are required to assess whether or not it is a Hycean world. In this work, we present JWST transmission spectroscopy of TOI-270~b and d, focusing specifically on the Hycean candidate TOI-270~d.

\section{Observations and data analysis} 
\label{sec:obs}

We conducted an atmospheric characterisation of the temperate sub-Neptune TOI-270~d using JWST transmission spectroscopy. The planet is the focus of multiple JWST programs with the JWST NIRISS, NIRSpec, and MIRI instruments (JWST GO Programs 2759, 3557, 4098). A transmission spectrum of the planet in the NIRSpec G395H band originally envisaged in multiple JWST programs, GO 3557 (PI: N. Madhusudhan) and GO 4098 (PI: B. Benneke), was allocated to the latter, with the data planned to be publicly accessible to the community. In this work, we investigated the resulting observation. The two primary transits of TOI-270~b and d occurred simultaneously during the observation between Oct 4, 2023 02:34:26 UTC and Oct 4, 2023 09:09:54 UTC, for a total exposure time of 5.3 hours. This rare event allows for transmission spectroscopy of both planets. The observation used the bright-object time series (BOTS) mode together with the F290LP filter, the SUB2048 subarray, and the NRSRAPID readout pattern. In this configuration, the spectrum is dispersed onto two different detectors, NRS1 and NRS2, spanning the wavelength ranges of 2.73-3.72 $\mu$m and 3.82-5.17 $\mu$m, respectively, with a gap in between. The spectral resolution of the G395H grating is R$\sim$2700, the highest resolution mode available with NIRSpec. The spectroscopic time-series observation is composed of 1763 integrations, with 11 groups per integration. 

We performed the data reduction and analysis according to \cite{Madhu_carbon2023}, using a variant of the \texttt{JExoRES} pipeline \citep{holmberg2023}. First, we use the JWST Science Calibration Pipeline \citep{Bushouse2020} for Stages 1 and 2, to calibrate the data and produce the count rate for each pixel and integration. During Stage 1, we perform an additional background subtraction step at the group level to mitigate 1/f noise. Second, we extracted the spectrum using the optimal extraction algorithm \citep{horne_optimal_1986}, while iteratively rejecting 5$\sigma$ outliers. Spectral channels with more than 20\% of the flux masked due to bad pixels or cosmic ray hits were discarded from further analysis.

\begin{figure}
        \includegraphics[width=0.5\textwidth]{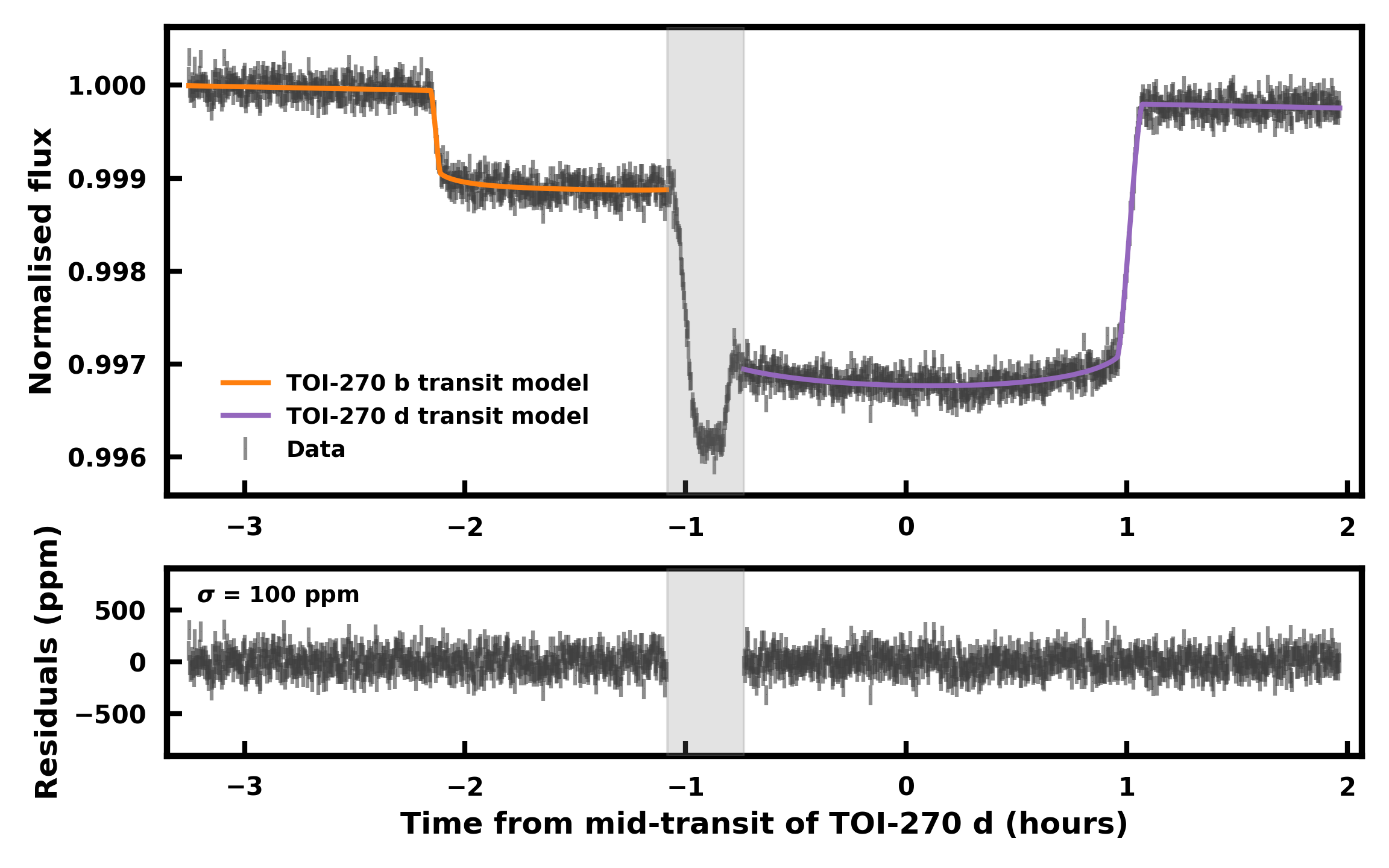}
  \vspace{-5mm}
    \caption{White light curve for the simultaneous transits of TOI-270 b and d observed with JWST NIRSpec G395H. The top panel shows the combined white light curve from NRS1 and NRS2, together with the best-fit models of the two transits shown in orange and purple. We mask the overlap region from the model fitting, indicated in grey. In both cases, we fit the transits using the out-of-transit flux on both sides. The bottom panel shows the residuals after subtracting the best-fit models. The standard deviation of the residuals is 100 ppm, corresponding to 1.2 times the expected noise level, similarly to other NIRSpec G395H observations.
    }
    \label{fig:wlc_nirspec} 
\end{figure}

We performed the light-curve analysis in three stages. In the first step we fit for the wavelength-independent system parameters, namely, the mid-transit time (T$_{0}$), the normalised semi-major axis ($a/R_*$), and the orbital inclination ($i$) using the integrated white light curve. Next, we binned the light curves into 256-pixel bins, corresponding to  $R \approx 20$, and fit the wavelength-dependent limb-darkening coefficients (LDCs). We discuss the effects of limb darkening on the transmission spectrum of TOI-270~d in Appendix \ref{app:LDCs}. Finally, using these system parameters and LDCs, we fit the light curves at the pixel level to obtain the transmission spectrum at high resolution. For the transit light-curve modelling, we used the aptly tested \texttt{batman} code \citep{kreidberg_batman_2015}. Furthermore, we assumed circular orbits for both planets and used the orbital periods $3.3601538$ days \citep{VanEylen2021} and $11.38014$ days \citep{Evans2023} for TOI-270 b and d, respectively. We adopted the two-parameter quadratic limb-darkening law using the parameterisation and priors by \cite{Kipping2013}. To model the baseline flux, we considered both linear and quadratic trends,  favouring a linear trend, as discussed in more detail in Appendix \ref{app:trend}. 

Figure \ref{fig:wlc_nirspec} shows the white light curve of the observation, obtained by integrating the light from both NRS1 and NRS2. We fit each transit separately by masking the integrations numbered 750 - 1480 for TOI-270 b and 385 - 865 for TOI-270 d. This  produces a 20-minute overlap window where both planets are transiting that we are disregarding from the analysis, as shown in grey in Figure \ref{fig:wlc_nirspec}. We also considered a case where we fit both transits simultaneously, as discussed in Appendix \ref{app:sim}. In either case, we did not consider the first five minutes of the observation due to a small settling ramp. We show the system parameters from the white light curve fitting in Appendix \ref{app:params}, obtained using \texttt{MultiNest} \citep{Feroz2009}. We use wide uniform priors for all parameters, except for $i$ and $a/R_*$, where we used normal priors with a mean and standard deviation given by \cite{VanEylen2021} for TOI-270 b and \cite{Evans2023} for TOI-270 d\footnote{We use the most conservative error estimate when unequal errors were given. For TOI-270 b, we calculate the priors of $a/R_*$ using the values and uncertainties of $a$ and $R_*$ given by \cite{VanEylen2021}.}. 

\begin{figure*}
    \includegraphics[width=\textwidth]{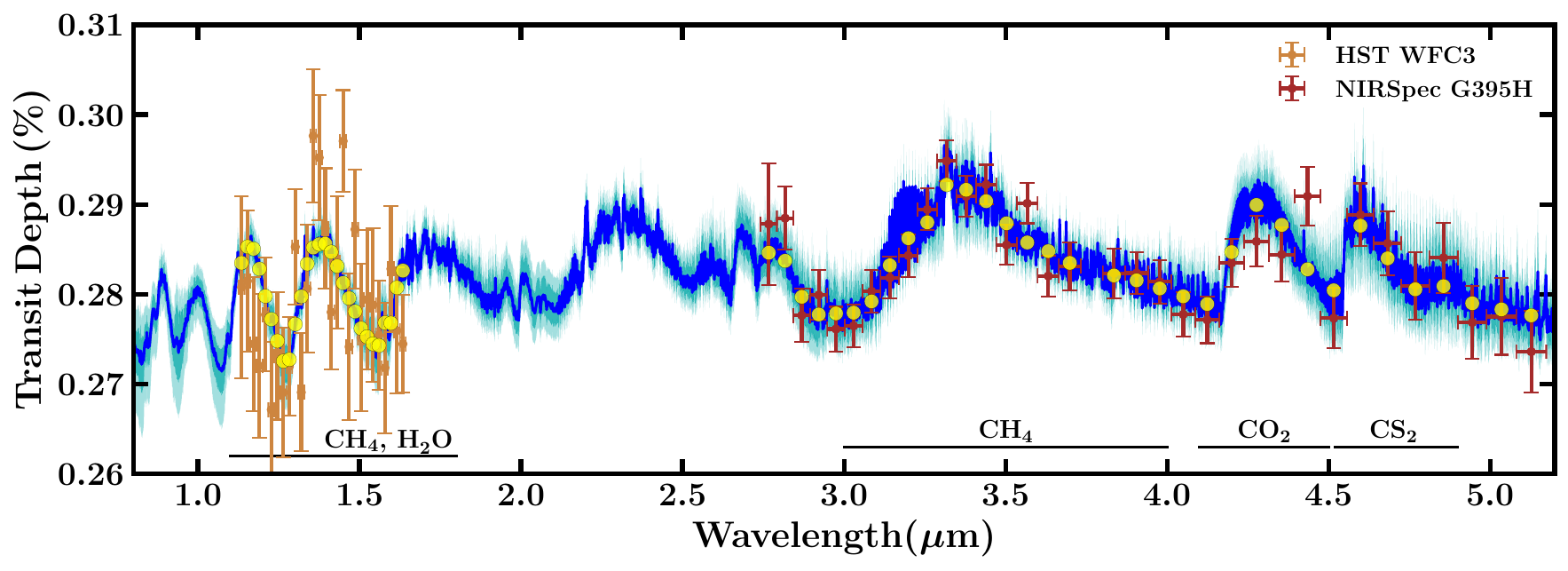}
     \vspace{-5mm}
    \caption{Transmission spectrum of TOI-270~d observed with JWST/NIRSpec G395H and HST/WFC3. The NIRSpec G395H spectrum between 2.7 - 5.2 $\mu$m, from this work, is shown in dark red, binned to $R \approx 55$ for visual clarity. The WFC3 spectrum is shown in orange and spans 1.1 - 1.6 $\mu$m, reported by \cite{Evans2023}. The retrieval is performed using the native resolution NIRSpec G395H spectrum ($R \sim 2700$) and the WFC3 spectrum. The NIRSpec spectrum is vertically offset by $-84$ ppm, corresponding to the median retrieved offset in the canonical one-offset case. The blue curve shows the median retrieved model spectrum, while the medium- and lighter-blue contours denote the 1$\sigma$ and 2$\sigma$ intervals, respectively. Yellow points correspond to the median spectrum binned to match the observations. The prominent molecules responsible for the features in different spectral regions are labelled.
    }
    \label{fig:spectrum} 
\end{figure*}

Finally, we performed the spectroscopic light fitting. We did this for each planet separately using the same masking as above and while fixing T$_{0}$, $i$, and $a/R_*$ to the values obtained from the white light curve analysis, as given in Table \ref{tab:wlc_params}. For our nominal spectrum, we bin the light curves into 256-pixel bins and fit these to obtain the wavelength-dependent LDCs ($u_1$ and $u_2$). We also consider a case assuming constant limb darkening, where we used the LDCs from the white light curve analysis \citep[similar to the \texttt{FIREFLy} pipeline; ][]{moran_high_2023, May2023}, as discussed in Appendix \ref{app:LDCs}. Equipped with empirical LDCs, we perform the fitting of the pixel-level light curves using the Levenberg–Marquardt algorithm to obtain the transit depths. This gives us two high-resolution transmission spectra, of TOI-270 b and d, covering the 2.73 - 5.17 $\mu$m wavelength range. We show the nominal spectrum of TOI-270 d in Figure \ref{fig:spectrum} together with our model fit, described in Section \ref{sec:retrieval}. We also reduce the data with the \texttt{Eureka!} pipeline \citep{Bell2022}, as shown in Appendix \ref{app:pipeline}. Finally, we present the spectrum of TOI-270~b in Appendix \ref{app:TOI_270b}, along with a comparative analysis with similar observations of K2-18~b \citep{Madhu_carbon2023} in Appendix \ref{app:K2_18b}.

\begin{figure*}
    \includegraphics[width=\textwidth]{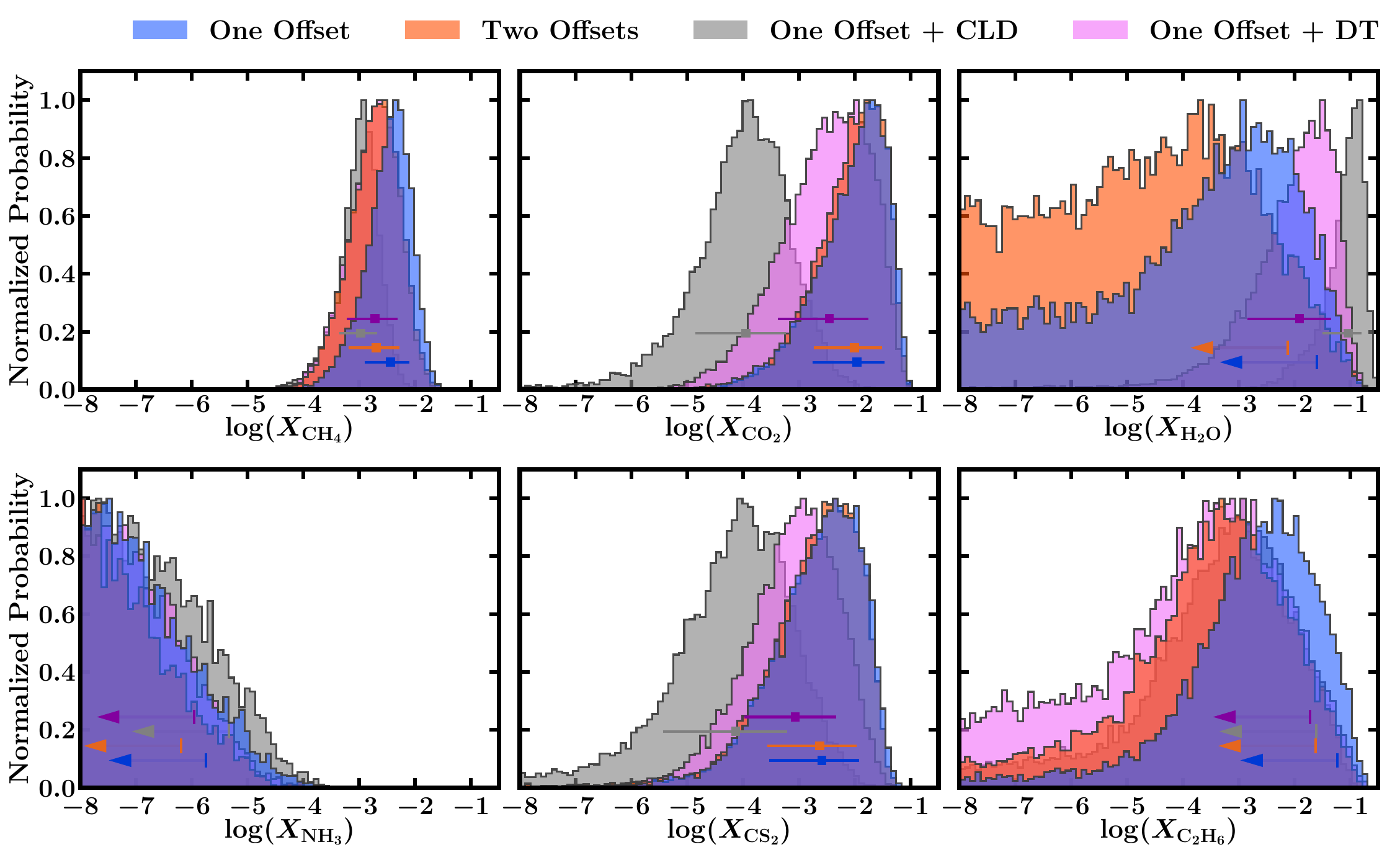}
     \vspace{-5mm}
    \caption{Retrieved posterior probability distributions for the mixing ratios of prominent molecules, shown for the four cases using WFC3 and NIRSpec data, as described in Section \ref{sec:retrieval}. The horizontal error bars denote the median and corresponding 1$\sigma$ interval for each distribution. The arrows in the case of H$_2$O, NH$_3$ and C$_2$H$_6$ indicate 95\% upper limits. We infer CH$_4$, CO$_2$, and CS$_2$ at 4.9$\sigma$, 3.6$\sigma,$ and 2.8$\sigma$ significance, respectively, in the canonical one-offset case with non-grey limb darkening. Abundance estimates and detection significance for these cases are shown in Tables~\ref{tab:abundances} and \ref{tab:retrieval_priors}.
    }
    \label{fig:posterior} 
\end{figure*}

\section{Atmospheric retrieval} \label{sec:retrieval}

We used the transmission spectrum of TOI-270~d from the current JWST NIRSpec observations, as well as data from the HST Wide Field Camera 3 \citep[WFC3;][]{Evans2023}, to retrieve the planet's atmospheric properties at the day-night terminator region. We employed the AURA retrieval code \citep{Pinhas2018} and adopted a methodology similar to previous studies of sub-Neptunes using simulated and real data from HST and/or JWST \citep[e.g.][]{Madhusudhan2020,Madhusudhan2021,Madhu_carbon2023, Welbanks2019,Constantinou2022, Evans2023}. We modelled the terminator region of the planet as a plane-parallel atmosphere in hydrostatic equilibrium, with a uniform chemical composition and assuming a hydrogen-dominated atmosphere. Within this framework, the chemical abundances, pressure-temperature ($P$-$T$) profile, and cloud-haze properties are treated as free parameters. In terms of chemical species, we incorporate molecular opacity originating from key CNO-bearing molecules typically expected in temperate H$_2$-rich atmospheres \citep{Pinhas2018,Welbanks2019,Constantinou2022}. For more details, see Appendix \ref{app:retrieval}.

To accommodate potential systematic effects, such as offsets between the NIRSpec and WFC3 spectra as well as between NIRSpec NRS1 and NRS2 \citep[as discussed in][]{Madhu_carbon2023}, we considered five retrieval scenarios. First, we considered our canonical case using data from both NIRSpec and WFC3 with an offset on NIRSpec relative to WFC3. Second, we considered two individual offsets for NRS1 and NRS2. We find that our canonical case, with no offset between NRS1 and NRS2, is preferred at 2.0$\sigma$ over the second scenario where such an offset is used. This suggests that the NIRSpec observation does not suffer from systematic offsets between the spectral baseline of NRS1 and NRS2. Nevertheless, here we discuss the constraints from both cases for completeness. Furthermore, we also explore scenarios involving constant limb darkening (CLD) and the simultaneous fitting of the dual transit (DT), as discussed in Appendices \ref{app:LDCs} and \ref{app:sim}, respectively. For  these two cases, we used an offset between NIRSpec and WFC3, as in our canonical case. Finally, we performed a retrieval with the NIRSpec spectrum alone. The results of these cases are shown in Table \ref{tab:abundances} and Table \ref{tab:retrieval_priors} and discussed below.

\begin{table*}
\small
\def\arraystretch{1.7}
\caption{Retrieved molecular abundances of prominent molecules in the atmosphere of TOI-270~d.
}
\vspace{-2mm}
\begin{tabular}{lccccccc}
\hline \hline
Cases & CH$_4$ & CO$_2$ & H$_2$O & NH$_3$ & CS$_2$ & C$_2$H$_6$ & CO\\
\hline
One offset & $-2.44_{-0.46}^{+0.34}$ (4.9$\sigma$) & $-1.96_{-0.79}^{+0.49}$ (3.6$\sigma$) & \textless$-1.56$ (1.6$\sigma$) & \textless$-5.75$ & $-2.59_{-0.95}^{+0.67}$ (2.8$\sigma$) & \textless$-1.23$ (2.3$\sigma$) & \textless$-1.63$ \\ 
 Two offsets & $-2.70_{-0.49}^{+0.42}$ (4.6$\sigma$) & $-2.01_{-0.73}^{+0.50}$ (3.8$\sigma$)  & \textless$-2.12$ (1.7$\sigma$) & \textless$-6.19$ & $-2.63_{-0.94}^{+0.66}$ (2.7$\sigma$) & \textless$-1.62$ (2.0$\sigma$)  & \textless$-2.01$  \\ 
  One offset + CLD & $-2.97_{-0.39}^{+0.30}$ (3.8$\sigma$) & $-3.95_{-0.90}^{+0.72}$ (2.9$\sigma$) & $-1.04_{-0.45}^{+0.24}$ (4.4$\sigma$) & \textless$-5.33$ & $-4.13_{-1.31}^{+0.92}$ (2.3$\sigma$) & \textless$-1.60$ (1.9$\sigma$) & \textless$-3.37$  \\
    One offset + DT & $-2.72_{-0.50}^{+0.41}$ (4.4$\sigma$) & $-2.46_{-0.92}^{+0.71}$ (3.9$\sigma$) & $-1.91_{-0.94}^{+0.57}$ (2.8$\sigma$) & \textless$-5.96$ & $-3.07_{-0.91}^{+0.74}$ (3.0$\sigma$) & \textless$-1.72$ (1.8$\sigma$) & \textless$-2.70$ \\ 
NIRSpec only & $-2.42_{-0.53}^{+0.40}$ & $-1.93_{-0.77}^{+0.49}$ & \textless$-1.99$ & \textless$-5.87$ & $-2.67_{-1.06}^{+0.73}$ & \textless$-1.16$ & \textless$-1.63$ \\ 
  \hline
\end{tabular}
\vspace{2mm}
\newline
\footnotesize{\textbf{Note.} These retrieval cases are described in Section~\ref{sec:retrieval}. The molecular abundances are shown as $\log_{10}$ of volume mixing ratios, with the detection significance (DS) in parentheses for the four cases using the full spectrum, including WFC3 and NIRSpec data. The retrieved median and 1$\sigma$ abundance estimates are given for CH$_4$, CO$_2$, CS$_2$, and (in two cases) H$_2$O. We give 95\% upper limits for the remaining molecules with low DS values. We note that the canonical one-offset case is preferred over the two-offset case at 2.0$\sigma$. Nevertheless, we show the abundance constraints for the two-offset case for completeness. The DS values shown here, obtained from Bayesian model comparisons using nested sampling, have typical uncertainties of $\sim$0.1$\sigma$ \citep{Madhu_carbon2023}. For reference, we also include the abundance constraints obtained with the NIRSpec data alone. All the retrieved properties for all the cases considered are shown in Table~\ref{tab:retrieval_priors}.}
\label{tab:abundances}
\end{table*}

\section{Results: Atmosphere of TOI-270~d}
\label{sec:270d}

The atmosphere of TOI-270~d was previously observed with HST WFC3 \citep{Evans2023}, reporting evidence of an H$_2$-rich atmosphere with H$_2$O absorption. Figure \ref{fig:spectrum}, shows the HST and JWST transmission spectra together with the retrieved spectral fit. In what follows, we outline the retrieved constraints on the atmospheric properties from the combined spectrum, including the present JWST NIRSpec spectrum (2.7–5.2 $\mu$m), and the HST WFC3 spectrum by \cite{Evans2023}, covering 1.1–1.6 $\mu$m. The retrieved constraints and detection significance are shown in Tables \ref{tab:abundances} and  \ref{tab:retrieval_priors}. 

\subsection{Prominent CNO molecules} \label{sec:cno}

Our atmospheric retrieval results support the inference of a H$_2$-rich atmosphere on TOI-270~d and provide valuable insights into the abundances of dominant CNO molecules. The spectrum reveals strong features of CH$_4$ and CO$_2$ in the atmosphere, as shown in Figure~\ref{fig:spectrum}, which are detected at 3.8-4.9$\sigma$ and 2.9-3.9$\sigma$, respectively, across the different retrieval cases (as shown in Table \ref{tab:abundances}). We computed the detection significance based on the Bayes factor, comparing the preference for a model fit to the data, while including a chemical species relative to the same model without it \citep{benneke2013, Pinhas2018}. Figure \ref{fig:posterior} shows the corresponding posterior probability distributions for several molecules of interest. The abundance estimates are summarised in Table~\ref{tab:abundances}. 

For our canonical one-offset retrieval, we obtained log volume mixing ratios of $\log$(X$_{\text{CH}_4}$) = $-2.44_{-0.46}^{+0.34}$ and $\log$(X$_{\text{CO}_2}$) = $-1.96_{-0.79}^{+0.49}$. These abundances are consistent with recent findings for K2-18 b \citep{Madhu_carbon2023}, which we discuss in Appendix \ref{app:K2_18b}. However, unlike K2-18~b, the effective partial transit in this work and the lack of precise short wavelength data with JWST NIRISS, means that the abundances are sensitive to the treatment of limb darkening. While in our canonical case we assumed a general non-grey treatment of limb darkening, we found that assuming constant limb darkening affects the abundances, resulting in $\log$(X$_{\text{CH}_4}$) = $-2.97_{-0.39}^{+0.30}$ and $\log$(X$_{\text{CO}_2}$) = $-3.95_{-0.90}^{+0.72}$, which we discuss in Appendix \ref{app:LDCs}. We also checked this for the recent NIRSpec G395H observation of K2-18~b \citep{Madhu_carbon2023} but found no significant difference in the abundances between the two treatments of limb darkening. Upcoming NIRISS observations of TOI-270~d will help to provide more accurate abundance estimates (JWST GO Programs 2759 and 4098). Nevertheless, we robustly detected CH$_4$ and CO$_2$ in the planet's atmosphere -- irrespective of the limb-darkening treatment.

We found only tentative evidence of H$_2$O, with the detection significance and abundance estimates varying across the retrieval cases. We do not robustly constrain H$_2$O given its limited spectral features in the NIRSpec G395H band and its degeneracy with CH$_4$ in the HST WFC3 band \citep{Blain2021, Bezard2022, Madhu_carbon2023}. Nevertheless, we did find some evidence of  H$_2$O, at 1.6-4.4$\sigma$ significance, depending on the data analysis approach. In the canonical retrieval, we found only a weak inference of H$_2$O (at 1.6$\sigma$), with a 95\% abundance upper limit of $-1.56$. This upper limit is consistent with the previous constraint from HST WFC3 by \cite{Evans2023}.  However, the detection significance of H$_2$O increases considerably for retrievals in the CLD (4.4$\sigma$) and DT (2.8$\sigma$) cases, as shown in Table \ref{tab:abundances}, with log-abundance of $-1.04_{-0.45}^{+0.24}$ and $-1.91_{-0.94}^{+0.57}$, respectively; also discussed in Appendix \ref{app:LDCs}. Therefore, more observations, for instance, with JWST NIRISS, as in the case of K2-18b \citep{Madhu_carbon2023}, would be required to conclusively constrain the presence and abundance of H$_2$O in the atmosphere of TOI-270~d.

We do not detect significant contributions from NH$_3$, CO or HCN. Nevertheless, we establish 95\% upper limits of $-5.75$, $-1.63$, and $-5.45$ for NH$_3$, CO, and HCN, respectively, in the canonical case, as shown in Table~\ref{tab:abundances}. The non-detections of these molecules are informative, given their anticipated strong spectral features and detectability within the wavelength range covered by WFC3 and NIRSpec G395H \citep{Madhusudhan2021, Constantinou2022}. Similarly to K2-18 b \citep{Madhu_carbon2023}, the present detections of CH$_4$ and CO$_2$ as well as the non-detection of NH$_3$, are consistent with the presence of a H$_2$O ocean under a thin H$_2$-rich atmosphere \citep{Hu2021, Madhusudhan2023}, making TOI-270~d a possible Hycean world.

\subsection{Other molecules} \label{sec:other}

We obtained notable constraints on two additional carbon-bearing molecules. We found 2-3$\sigma$ evidence of carbon disulfide  (CS$_2$) across the retrievals. In particular, CS$_2$ is of great interest since it has been predicted to be a detectable biomarker in Hycean atmospheres \citep{Madhusudhan2021} as well as for rocky exoplanets with H$_2$-rich atmospheres \citep{seager2013b}, albeit with sources including abiotic mechanisms (discussed in Section \ref{sec:discussion}). For the canonical retrieval, we inferred an abundance of $\log$(X$_{\text{CS}_2}$) = $-2.59_{-0.95}^{+0.67}$, which is consistent with the other retrieval cases within 1$\sigma$, as shown in Table~\ref{tab:abundances}. The detection significance of CS$_2$ is 2.7-3.0$\sigma$ for three of the four cases using WFC3 and NIRSpec data (see Table~\ref{tab:abundances}), with the CLD case showing a lower significance of 2.3$\sigma$. 

Additionally, we found potential hints of C$_2$H$_6$, at a lower significance of 1.8-2.3$\sigma$ across the retrievals. C$_2$H$_6$ is known to be a photochemical byproduct of reactions involving CH$_4$ and other organic molecules, including several biogenic gases \citep{domagal-goldman2011, catling2018, schwieterman2018}. However, we note that both C$_2$H$_6$ and CH$_4$ have strong spectral features in the $\sim$3.3-3.5 $\mu$m range (shown in Figure \ref{fig:contribution}), making it difficult to reliably infer C$_2$H$_6$ with the current data. Given the low detection significance, we only consider an upper limit for the abundance of C$_2$H$_6$. The 95\% upper limit for its log mixing ratio is between $-1.72$ and $-1.23$ across the retrievals, as shown in Table~\ref{tab:abundances}. 

Finally, we note that for both CS$_2$ and C$_2$H$_6$ the retrieved abundances are anomalously high compared to theoretical expectations \citep{domagal-goldman2011, seager2013b, seager2016}. More observations are required to robustly constrain the presence and abundances of both molecules.

\subsection{Clouds, hazes and temperature structure} \label{sec:clouds}

The observed transmission spectrum does not provide strong constraints on clouds or hazes, similarly to the previous HST observation \citep{Evans2023}, due to limited data at shorter wavelengths. The constraints on the cloud-top pressure due to grey clouds are weak and mostly lie below the observable photosphere (e.g. cloud top pressures $\geq $100 mbar). We find that a model with no clouds or hazes is somewhat favoured at 1.6$\sigma$ in the canonical case. However, this does not preclude the possibility of clouds or hazes deeper in the atmosphere. The weak constraints on the cloud and haze parameters are shown in Table \ref{tab:retrieval_priors}. We note that not including clouds or hazes does not significantly affect the abundance constraints. Upcoming observations with JWST NIRISS at shorter wavelengths will be able to further constrain the presence of clouds and hazes (JWST GO Programs 2759 and 4098).

The observations provide nominal constraints on the temperature within the planetary photosphere. Specifically, we find that the temperature at 10 mbar is $305^{+101}_{-93}$ K in the canonical one-offset case (as detailed in Table \ref{tab:retrieval_priors}). We also compared the effect of assuming an isothermal $P$-$T$ profile in the canonical case, instead of the nominal six-parameter profile, but found no preference for either model. The abundance constraints are consistent between the two assumptions. It is worth noting that temperature determination through transmission spectroscopy tends to be less sensitive compared to emission spectroscopy \citep{Madhusudhan2014}. Nevertheless, the retrieved temperature range is consistent with the possibility of H$_2$O in the observable atmosphere. We discuss the implications of these findings and compare them with K2-18~b in Appendix \ref{app:K2_18b}. 

In the context of a H$_2$-rich Hycean atmosphere, it is important to recognise that such an atmosphere can induce a substantial greenhouse effect, leading to the warming of the ocean surface. Therefore, the presence of clouds and/or hazes plays a pivotal role in cooling the atmosphere \citep{Madhusudhan2020, Madhusudhan2021, Madhusudhan2023, Piette2020}, and may result in more moderate conditions at the ocean surface compared to what would be expected by models that do not consider clouds or hazes \citep[e.g.][]{Scheucher2020, Innes2023}.

\subsection{Comparative characterisation between TOI-270 b and d} \label{sec:bd}

The innermost planet, TOI-270~b, is significantly smaller and hotter than the Hycean candidate TOI-270~d. The planet is a super-Earth with a radius of 1.2-1.3 $R_\oplus$ with a zero-albedo equilibrium temperature of $581\pm14$ K \citep{VanEylen2021}. The present JWST observations provide a unique opportunity to characterise two sub-Neptunes in the same system. Our  transmission spectrum of TOI-270~b is shown in Figure \ref{fig:T270b}. Because of its size, we do not expect strong atmospheric features as those seen in the case of TOI-270~d. Nevertheless, the spectrum of TOI-270~b shows potential spectral features that are consistent with an H$_2$-rich atmosphere, deviating from a featureless spectrum at 3.3-3.5$\sigma$ (as discussed in Appendix \ref{app:TOI_270b}). Using the constraints on stellar heterogeneities from the spectrum of TOI-270~d, we find that a H$_2$-rich atmosphere is preferred over a contaminated flat spectrum at 2.7$\sigma$. However, given the low signal-to-noise ratio (S/N), more observations are required to robustly distinguish these features from the effects of stellar heterogeneities and potential systematics.

\section{Summary and discussion} \label{sec:discussion}

 We investigated JWST observations of the transmission spectra of two sub-Neptunes simultaneously transiting the nearby M dwarf TOI-270. The spectra were obtained using the JWST NIRSpec G395H instrument in the 2.7-5.2 $\mu$m wavelength range. The planets include a candidate Hycean world, TOI-270~d, and a super-Earth, TOI-270~b. In this paper, we report our atmospheric characterisation of TOI-270~d using the JWST spectrum, combined with a previous HST WFC3 spectrum in the 1.1-1.6 $\mu$m range. 

 \subsection{Chemical detections}
 
 We robustly detected CH$_4$ and CO$_2$ in the atmosphere of TOI-270~d, with 3.8-4.9$\sigma$ and 2.9-3.9$\sigma$ confidence, respectively, and with volume mixing ratios of $\sim$0.1-1\%. However, these abundances differ somewhat depending on the limb-darkening assumptions. We also found tentative evidence of H$_2$O, with varying confidence levels and abundance estimates, depending on the data analysis approach, but which are consistent with the previous inference using HST data \citep{Evans2023}. On the other hand, our robust detection and abundance estimate of CH$_4$ are particularly significant considering its previous non-detection using the HST WFC3 spectrum alone \citep{Evans2023}. This is similar to the case of K2-18 b, where the recent JWST observations \citep{Madhu_carbon2023} resolved the degeneracy between CH$_4$ and H$_2$O that had originally been inferred from the HST WFC3 data \citep{Blain2021, Bezard2022}. We did not detect NH$_3$, CO, or HCN in the atmosphere, establishing upper limits on their abundances. 
 
 Additionally, we found evidence for the presence of CS$_2$ at a detection significance of 2.3-3.0$\sigma$, and potential hints of C$_2$H$_6$ at 1.8-2.3$\sigma$. Together with the present data, upcoming JWST observations of TOI-270~d, especially at shorter wavelengths, will be able to provide more accurate and precise abundance constraints \citep{Constantinou2023}. For TOI-270~b, we find some evidence of a H$_2$-rich atmosphere; however, more observations are needed to robustly confirm these results and rule out systematics.

\subsection{TOI-270~d: A potential Hycean world}

TOI-270~d has been predicted to be a candidate Hycean world \citep{Madhusudhan2021}, with a potentially habitable ocean underneath a H$_2$-rich atmosphere. The planet's atmosphere was first observed with HST \citep{Evans2023}, showing evidence of H$_2$O in a H$_2$-rich atmosphere. The present NIRSpec spectrum, leading to detections of CH$_4$ and CO$_2$ at high significance, as well as the non-detection of  NH$_3$, support the interpretation of TOI-270~d as a Hycean planet. Similarly to what was observed for K2-18~b \citep{Madhu_carbon2023}, this composition cannot be explained by a Neptune-like deep H$_2$-rich atmosphere, but it is consistent with that of a warm Hycean world, as predicted in \cite{Madhusudhan2023}. In particular, the inferred temperature and the tentative evidence of H$_2$O, at 1.6-4.4$\sigma$ suggest that the day-night terminator region is warm enough for H$_2$O to remain gaseous at the pressures probed in transmission spectroscopy instead of condensing out. We note that the generally higher photospheric temperature at the terminator of TOI-270~d, as compared to K2-18~b, indicates that the dayside surface may be too hot for a liquid ocean, potentially reaching supercritical temperatures in the absence of a strong albedo from clouds or hazes \citep{Piette2020, Scheucher2020, Pierrehumbert2023, Innes2023}. However, the retrieved temperature of TOI-270~d is cool enough so that if the planet is tidally locked, there could still be conditions allowing for a habitable liquid ocean on the night side, which would correspond to a dark Hycean world \citep{Madhusudhan2021}. 

We also find evidence of CS$_2$ in the atmosphere of TOI-270~d. This is of particular interest given that CS$_2$ is a predicted biosignature gas \citep{domagal-goldman2011, Seager2013, Madhusudhan2021}. Additionally, C$_2$H$_6$, for which we find potential hints, is a predicted indicator of photochemical processes involving CH$_4$ and other organic molecules, including several gases of biological origin \citep{domagal-goldman2011, catling2018, schwieterman2018}. However, it is important to note that, in comparison to DMS that was nominally inferred for K2-18~b \citep{Madhu_carbon2023}, these molecules have alternative abiogenic sources \citep{Rushdi2005, domagal-goldman2011, catling2018, schwieterman2018}. Upcoming observations could allow more robust inferences of CS$_2$ and C$_2$H$_6$ in the atmosphere of TOI-270~d at a higher significance -- if they are indeed present. 

In conclusion, the current JWST transmission spectrum of TOI-270~d offers rich insights into the atmospheric composition of the planet. The planet stands out as a promising Hycean candidate, consistent with its initial predictions as a world with the potential for habitable oceans beneath a H$_2$-rich atmosphere. The robust detection of CH$_4$ and CO$_2$ and the non-detection of NH$_3$ strengthen the case for its classification as a Hycean world. Together with the recent work on K2-18~b \citep{Madhu_carbon2023}, these findings demonstrate the beginning of a new era in atmospheric characterisations of habitable exoplanets with JWST.\\

 \vspace{-3mm}
\begin{acknowledgements}
This work is based on observations made with the NASA/ESA/CSA James Webb Space Telescope as part of Cycle 2 GO Program 4098 (PI: B. Benneke). This work is supported by research grants to N.M. from the UK Research and Innovation (UKRI) Frontier Grant (EP/X025179/1), the MERAC Foundation, Switzerland, and the UK Science and Technology Facilities Council (STFC) Center for Doctoral Training (CDT) in Data Intensive Science at the University of Cambridge (STFC grant No. ST/P006787/1). N.M. and M.H. acknowledge support from STFC and the MERAC Foundation toward the doctoral studies of M.H. We thank the reviewer for their detailed review and valuable comments on the manuscript.

This work was performed using resources provided by the Cambridge Service for Data Driven Discovery operated by the University of Cambridge Research Computing Service (\url{www.csd3.cam.ac.uk}), provided by Dell EMC and Intel using Tier-2 funding from the Engineering and Physical Sciences Research Council (capital grant EP/P020259/1), and DiRAC funding from STFC (\url{www.dirac.ac.uk}).\\[4pt]

{\it Author contributions. } N.M. conceived and planned the project. M.H. conducted the data reduction and light-curve analysis. N.M. conducted the atmospheric retrieval analysis and interpretation. M.H. led the writing of the manuscript with guidance and contributions from N.M.\\[4pt]

{\it Data availability.} The JWST NIRSpec transmission spectra of TOI-270~d reported in this work following different data analysis approaches and used in the atmospheric retrievals are available on the Open Science Framework at \url{https://osf.io/8fu36}.

\end{acknowledgements}

\bibliography{ms.bib, references.bib}
\bibliographystyle{aa}

\begin{appendix}

\section{System parameters} \label{app:params}
Here, we estimate the system parameters and planet radii from the white light curve analysis of the JWST NIRSpec G395H observations of TOI-270 b and d. We fit each transit separately, as described in Section \ref{sec:obs}. The estimated parameters are given in Table \ref{tab:wlc_params}, where $u_1$ and $u_2$ represent the two quadratic limb-darkening coefficients. 

\begin{table}[h]
\def\arraystretch{1.7}
\caption{Estimated parameters from the white light curve analysis of the JWST NIRSpec G395H observation of TOI-270 b and d.}
\vspace{-2mm}
\begin{tabular}{lcccc}
\hline \hline
Parameter & TOI-270 b & TOI-270 d \\ \hline
T$_0$ (BJD) & $60221.24113_{-0.00095}^{+0.00086}$ &  $60221.30143_{-0.00043}^{+0.00050}$ \\ 
$i$ ($^{\circ}$) & $89.39_{-0.29}^{+0.32}$ & $89.746_{-0.069}^{+0.080}$  \\ 
$a / R_*$ & $18.34_{-0.45}^{+0.42}$ & $42.13_{-0.35}^{+0.37}$ \\ 
$R_\mathrm{p} / R_*$ & $0.03142_{-0.00015}^{+0.00014}$ & $0.054072_{-0.000081}^{+0.000091}$\\
$u_1$ & $0.056_{-0.039}^{+0.058}$ & $0.130_{-0.025}^{+0.024}$  \\ 
$u_2$ & $0.163_{-0.092}^{+0.077}$ & $0.054_{-0.043}^{+0.042}$ \\ 
$R_\mathrm{p}$ ($R_\oplus$) & $1.302 \pm 0.028$ & $2.241 \pm 0.047$ \\ \hline
\end{tabular}
\vspace{2mm}
\newline
\footnotesize{\textbf{Note.} The orbital periods are held fixed at $3.3601538$ days \citep{VanEylen2021} and $11.38014$ days \citep{Evans2023} for TOI-270 b and TOI-270 d, respectively. We use a stellar radius of $0.380 \pm 0.008$ $R_\odot$ \citep{Kaye2022} to estimate the planet radii. Also, note that the mid-transit time is given in units of BJD - 2400000.5 days.}
\label{tab:wlc_params}
\end{table}

\section{Light-curve detrending and systematics} \label{app:trend}

Recent transit observations with the NIRSpec G395H instrument have revealed systematic trends in the baseline flux \citep{Alderson2023, Espinoza2023, moran_high_2023, Madhu_carbon2023}. Typically, these trends are modelled using low-order polynomials. However, the choice of polynomial order can impact the measured transit depths, as discussed in \cite{Madhu_carbon2023}. Therefore, we investigated the influence of detrending on the current transmission spectra of TOI-270 b and d, considering both linear and quadratic detrending approaches.

For the quadratic detrending procedure, we adopted the approach outlined in \cite{moran_high_2023} and \cite{Madhu_carbon2023}. In this method, the quadratic-trend parameter is held fixed during the spectroscopic light curve analysis to the value obtained from the detector-specific white light curve. To achieve this, we performed separate fits to the transit of TOI-270 d for both NRS1 and NRS2 white light curves (masking the transit of planet b). These fits yield the quadratic-trend parameters $5.7\pm3.6$ and $4.3\pm4.3$ ppm/hour$^2$ for NRS1 and NRS2, respectively. Subsequently, we employed these parameter values to fit the spectroscopic light curves and derive the transmission spectra of both planets.

In line with the findings reported in \cite{Madhu_carbon2023}, we find that the primary distinction between the linear- and quadratic-trend assumptions on the transmission spectra is the presence of a constant offset, affecting NRS1 and NRS2 differently. Specifically, for TOI-270~d, we quantify the offsets between the linear and quadratic cases to be approximately 19 and 15 ppm for NRS1 and NRS2, respectively. Furthermore, we observe that deviations beyond a constant offset do not exceed a few ppm, which corresponds to a tenth of the transit depth uncertainty at R = 100.

In light of these findings, we opted to utilise the linear-trend spectra in our work. Additionally, we considered a retrieval scenario in which we account for the potential presence of an offset between NRS1 and NRS2, similar to the approach in \cite{Madhu_carbon2023}, to capture any associated effects. However (as discussed in Section \ref{sec:retrieval}), we did not find evidence of an offset between NRS1 and NRS2 given that a model with no offset between the two detectors is preferred over a model with such an offset at 2.0$\sigma$. For this reason, we conclude that the present NIRSpec observation is likely not affected by a systematic offset between the spectral baseline of NRS1 and NRS2.

\section{Comparing different pipelines} \label{app:pipeline}

In what follows, we compare the transmission spectrum of TOI-270~d from \texttt{JExoRES} and \texttt{Eureka!}. We also offer general remarks on the effectiveness of comparing multiple data reduction pipelines as a way to test the robustness of the analysis.

\subsection{Comparison with \texttt{Eureka!}} \label{app:Eureka}

\begin{figure*}
        \includegraphics[width=\textwidth]{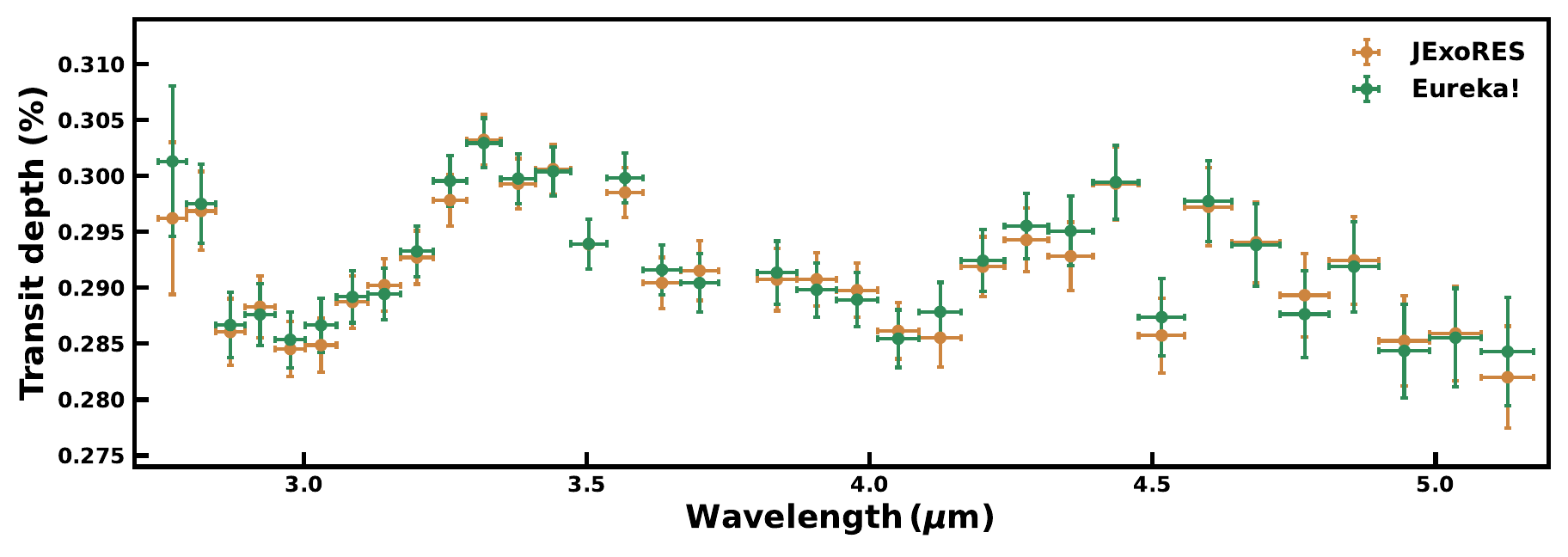}
 \vspace{-5mm}
    \caption{Comparison of the \texttt{JExoRES} and \texttt{Eureka!} transmission spectra of TOI-270~d, binned to $R \approx 55$ for visual clarity. The \texttt{JExoRES} and \texttt{Eureka!} spectra are shown in orange and green, respectively. Overall, we find a good agreement between the two data reductions.
    }
    \label{fig:Eureka_comparison} 
\end{figure*}

We compared the TOI-270~d transmission spectrum in this work to that obtained from the \texttt{Eureka!} data reduction and analysis pipeline \citep{Bell2022}, following the steps and parameters outlined in \cite{May2023}, unless otherwise stated below. Similar to \texttt{JExoRES}, \texttt{Eureka!} uses the \texttt{jwst} pipeline \citep{bushouse_2022} for Stages 1 and 2, while also implementing a custom group-level background subtraction to correct for the effects of 1/f noise. We used the jump detection step with a rejection threshold of 5$\sigma$. In Stage 3 of \texttt{Eureka!}, we performed a second round of background subtraction (not in \texttt{JExoRES}) and extracted the spectrum according to \cite{horne_optimal_1986}, using the same outlier rejection thresholds for both detectors as the NRS1 reduction in \cite{May2023}. We generated the light curves at the pixel level in Stage 4 and manually disregarded a few spectral channels with much higher noise than nearby channels. We also used a rolling median to reject 5$\sigma$ outliers. In Stage 5, we modelled the spectroscopic light curves with the \texttt{batman} transit model \citep{kreidberg_batman_2015} together with a linear slope for the baseline flux, in line with \texttt{JExoRES}. We fixed the system parameters to the values in Table \ref{tab:wlc_params} and used the same set of quadratic LDCs as for the \texttt{JExoRES} data reduction. We also masked the same number of integrations (as described in Section \ref{sec:obs}) so that our results would not be affected by the transit of TOI-270~b. For each light curve, we fit for $R_\mathrm{p} / R_*$, the two linear trend parameters, and the uncertainty scaling parameter. We obtained the transit depth from the least-squares best-fit and the uncertainties from \texttt{emcee} \citep{foreman-mackey_emcee_2013}. Overall, as shown in Figure \ref{fig:Eureka_comparison}, we see a good agreement between the two different data reductions, as expected given the similar methods used. However, we note that there are some differences between the two data reductions, mainly regarding the background subtraction, optimal extraction profiles, as well as the methods and thresholds used to flag and mask outliers.

\subsection{Regarding robustness checks with multiple pipelines}

Several works have claimed to have achieved a good agreement between different data reductions of JWST transmission spectroscopy observations \citep[e.g.][]{Ahrer2023, Alderson2023, Feinstein2023, Rustamkulov2023}. However, given the sensitivity of atmospheric retrievals to the high-precision data from JWST\footnote{Atmospheric retrievals are designed to be maximally sensitive to the data in order to extract as much information as possible.}, small differences in the planet's spectrum from different data reductions can affect abundance estimates and the detectability of less prominent species, especially when  shorter wavelength data \citep[$\lesssim 1.5$ $\mu$m;][]{Constantinou2023} are lacking. Because of this issue, it is arguably desirable to perform multiple data reductions as a way to test the robustness of the analysis \cite{lustig-yaeger_jwst_2023, moran_high_2023, Radica2023, Lim2023, Grant2023, May2023, Fournier2023, Bell2023}. However, it may be argued that this approach does not necessarily ensure the reliability of the analysis. First, it is important to note that a pipeline itself is only as good as the underlying assumptions and methods used to reduce the data. For example, our comparison between \texttt{JExoRES} and \texttt{Eureka!} demonstrates intercomparisons between implementations of what are essentially two very similar data reductions. Although cross-checks are useful, they do not necessarily qualify as an argument for the overall robustness of the data reduction. Instead, it is important to ensure that within any given pipeline, multiple assumptions and approaches for reducing and analysing the data are tested thoroughly, as we have aimed to do in this work.

Second, care must be taken when the transmission spectrum is found to vary between different data reduction assumptions, especially when these have not been verified. For example, \cite{Grant2023} found that different approaches to the correction of detector non-linearity and the Brighter-Fatter effect \citep{Argyriou2023} in the \texttt{ExoTiC-MIRI} and \texttt{Eureka!} pipelines led to differences in the MIRI transmission spectrum of WASP-17~b. In that study, \cite{Grant2023} chose to rely on only one of the data reductions for their conclusions, leading to the claim of a 2.6$\sigma$ detection of quartz clouds. However, their conclusions might have been different if both reduction approaches were considered on equal footing. In the present case, as demonstrated by the intercomparison between \texttt{JExoRES} and \texttt{Eureka!}, the transmission spectrum of TOI-270~d is stable against minor differences in the background subtraction, spectrum extraction and the treatment of outliers. Nevertheless, we have investigated several effects that have been shown to be important in the data reduction and analyses of such data, including detector offsets, baseline trends and limb-darkening treatments, as discussed in Appendix~\ref{app:LDCs}.

\section{Stellar limb darkening} \label{app:LDCs}

\begin{figure}
        \includegraphics[width=0.493\textwidth]{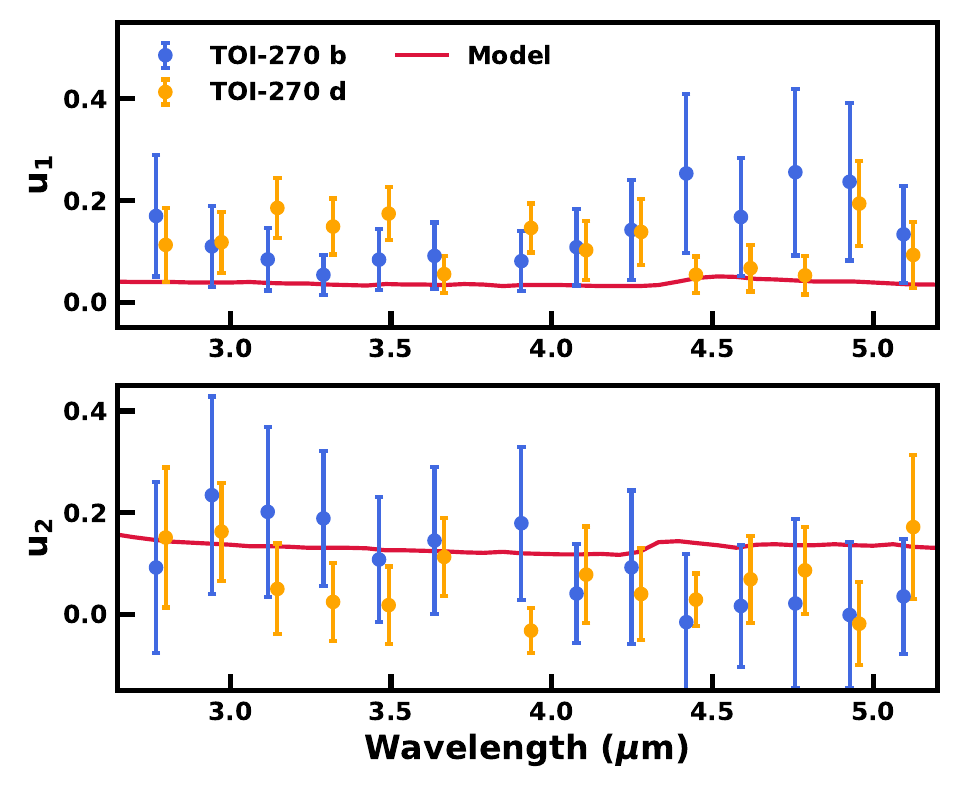}
  \vspace{-5mm}
    \caption{Empirical quadratic LDCs of TOI-270. The fitted LDCs from the transit of TOI-270 b and d are shown in blue and orange, respectively. The LDCs obtained from the transit of TOI-270~d are shown with a small wavelength offset for clarity. The red curves show the quadratic LDCs predicted by the ATLAS9 stellar atmosphere models, obtained from ExoCTK. As can be seen, we find some potential differences between the empirical and theoretical LDCs. 
    }
    \label{fig:LDCs} 
\end{figure}

\begin{figure*}
        \includegraphics[width=\textwidth]{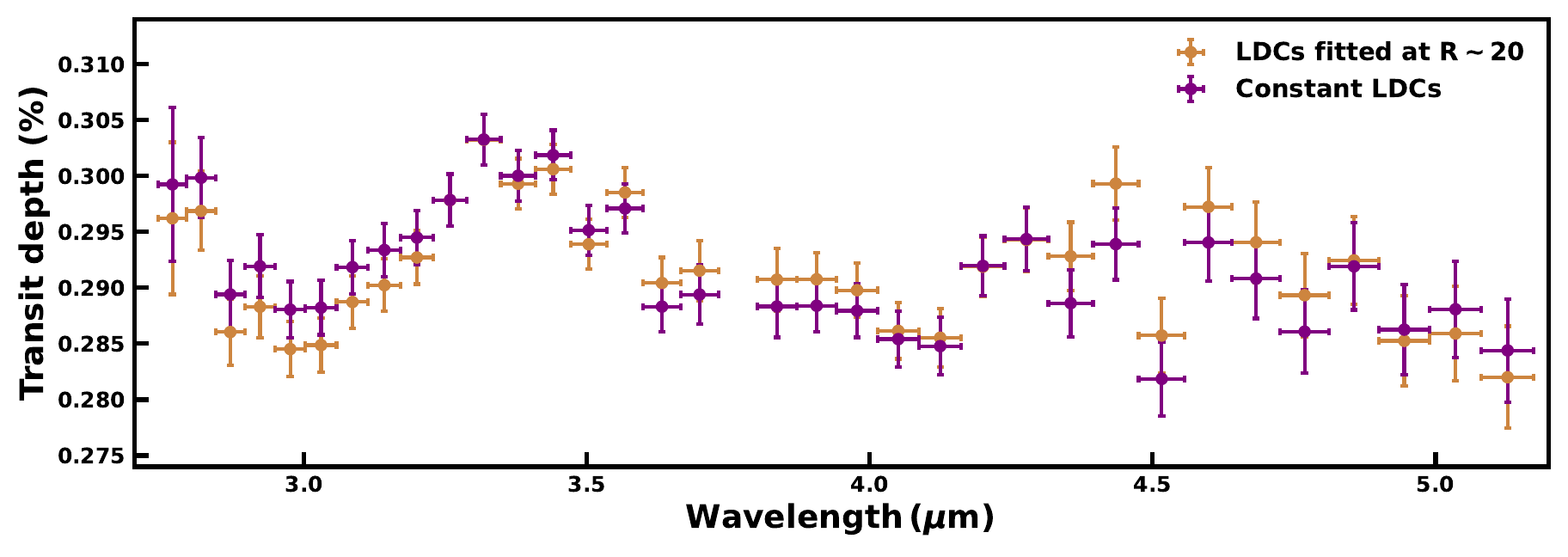}
  \vspace{-5mm}
    \caption{Effect of stellar limb-darkening assumptions on the transmission spectrum of TOI-270~d. The nominal spectrum with wavelength-dependent LDCs fitted at $R \approx 20$ is shown in orange. The spectrum assuming a constant (grey) limb darkening, using the LDCs from the white light curve fit, is shown in purple.}
    \label{fig:LDCs_comparison} 
\end{figure*}

The treatment of stellar limb darkening is known to affect the derived transit depth and orbital parameters \citep[e.g.][]{Espinoza2016, Maxted2018}. This effect may be particularly important for the present observation given the simultaneous transit of TOI-270~b and d, effectively resulting in partial transits after masking, which have made it more difficult to determine the transit properties. For these reasons, we investigated the effects of different limb-darkening assumptions on the transmission spectrum of TOI-270~d. 

Instead of relying on stellar atmosphere models, which can be inaccurate for stars cooler than about 5000 K \citep{Patel2022}, we chose to fit for the stellar limb darkening parameters. As shown in Figure \ref{fig:LDCs}, we find some potential disagreement when comparing the empirical quadratic LDCs to the theoretical values from the ATLAS9 1D stellar atmosphere models \citep{Castelli2023}, obtained from ExoCTK \citep{matthew_bourque_2021}. We note that we could not use the Stagger 3D stellar atmosphere models by \cite{Magic2015}, commonly used to derive theoretical LDCs \citep[e.g.][]{lustig-yaeger_jwst_2023, moran2020, May2023}, since the grid does not contain effective temperatures below 4000 K. To first order, there appears to be a shift in the model LDCs compared to the fitted LDCs of TOI-270, similar to the case of WASP-39 \citep{Rustamkulov2023}. To quantify this disagreement, we compare the Bayesian evidence from the white light curve fitting in the nominal case, where we fit the LDCs, and when we fix the LDCs to the model values ($u_1 = 0.037$ and $u_2 = 0.13$). We add up the (log) evidence from both transit fits and compute the difference between the two cases. This results in a 4.4$\sigma$ preference for the fitted LDCs compared to the fixed model LDCs. Therefore, we may risk biasing the transmission spectrum using the model LDCs given that these do not appear to be a good match to the data.

Given the approximately constant theoretical limb darkening of TOI-270 in the G395H wavelength range, we derived an alternative transmission spectrum of TOI-270~d, assuming that the limb darkening is wavelength-independent. For this, we use the LDCs from the white light curve fit of TOI-270~d, as shown in Table \ref{tab:wlc_params}. Figure \ref{fig:LDCs_comparison} illustrates the difference between the two assumptions. The largest differences can be seen below 3.5 $\mu$m and around 4.5 $\mu$m, near the CO$_2$ feature. The abundance estimates and detection significances of this alternative transmission spectrum are given in Table \ref{tab:abundances}, obtained using the same atmospheric retrieval configuration as in the canonical one-offset case presented in Section \ref{sec:retrieval}. We find that the difference in limb-darkening has two main effects on the interpretation of the spectrum of TOI-270~d. First, the elevated spectral baseline below 3.5 $\mu$m results in a higher H$_2$O abundance for the constant limb-darkening case. In fact, we detect H$_2$O at 4.4$\sigma$ assuming wavelength-independent limb darkening. However, more observations are needed to reliably constrain the presence and abundance of H$_2$O. Second, the shallower spectral features at around 4.5 $\mu$m lead to a decrease in the abundances of CO$_2$ and CS$_2$. Despite these differences, we robustly detected the presence of CH$_4$ and CO$_2$ in the planet's atmosphere.

For comparison, we also investigated the effects of limb darkening on the NIRSpec G395H observation of the sub-Neptune K2-18~b \citep[JWST GO Program 2772, PI: N. Madhusudhan; ][]{Madhu_carbon2023}, given the similarity to TOI-270~d. In this case, we find that the choice of limb-darkening treatment (i.e. wavelength-dependent or not) had a negligible effect on the inferred atmospheric properties of the planet. The higher level of sensitivity to the treatment of limb darkening for the present observation of TOI-270~d (compared to K2-18~b) may be attributed to the lack of precise short wavelength data (coming from NIRISS in the case of K2-18~b) and the absence of data around ingress in the present case (when masking the transit of TOI-270~b). Future JWST observations of TOI-270~d covering a larger wavelength span will be able to better constrain the atmospheric properties of the planet. Nevertheless, the detections of CH$_4$ and CO$_2$ and the non-detection of NH$_3$ are independent of the treatment of limb darkening.

\section{Simultaneous transit modelling} \label{app:sim}

\begin{figure}
        \includegraphics[width=0.5\textwidth]{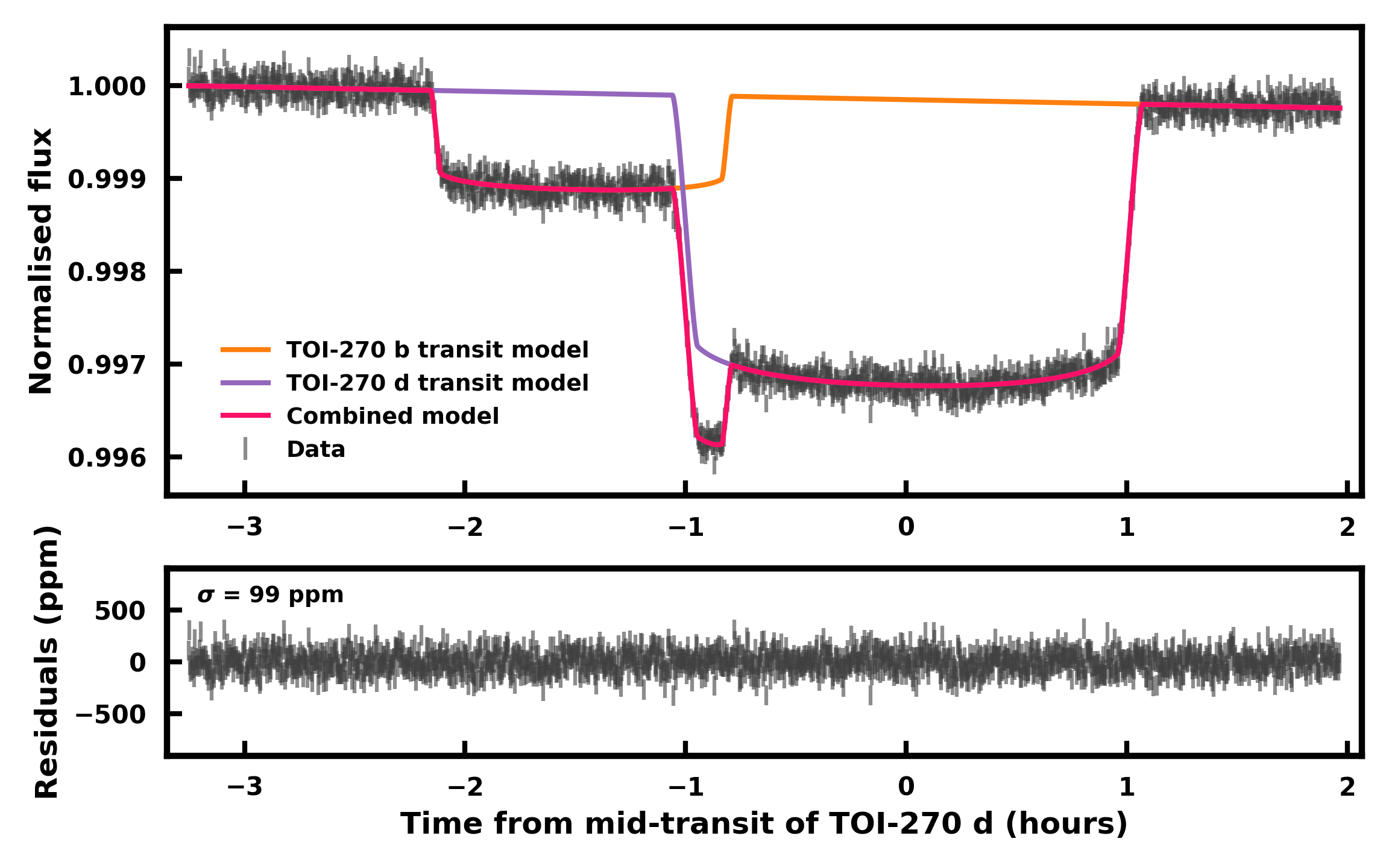}
  \vspace{-5mm}
    \caption{Simultaneous model fit of the transits of TOI-270~b and d. The top panel shows the combined white light curve from NRS1 and NRS2, together with the best-fit models. The bottom panel shows the residuals after subtracting the best-fit combined model. The standard deviation of the residuals is 99 ppm.
    }
    \label{fig:sim} 
\end{figure}

We also considered an additional case where we simultaneously fit the transits of both TOI-270~b and d. We modelled the combined light curve from the dual transit (DT) as $F_\mathrm{poly}(t) (F_\mathrm{b}(t) + F_\mathrm{d}(t) - 1)$, where $F_\mathrm{poly}(t)$ is a linear polynomial describing the out-of-transit flux, and $F_\mathrm{b}(t)$ and $F_\mathrm{d}(t)$ are the transit light curves of TOI-270~b and d, respectively, modelled using \texttt{batman} \citep{kreidberg_batman_2015}. Figure \ref{fig:sim} shows the white light curve together with the combined model fit, illustrating that the model provides a good fit to the data. We do not find evidence for the planets occulting each other during the transits. The inferred system parameters, LDCs (now the same for both planets), and transit depths from the simultaneous model fit of the white light curve are all within $\sim$1$\sigma$ of the values determined from the separate fits of the two transits, given in Table \ref{tab:wlc_params}. To generate the spectra, we fit the spectroscopic light curves as described in Section \ref{sec:obs}, assuming non-grey limb darkening as for our nominal spectra and use the present system parameters, while only masking the first five minutes, as both transits were now being fitted together. The retrieved  atmospheric properties of TOI-270~d in this case are shown in Figure \ref{fig:posterior} and Tables \ref{tab:abundances} and \ref{tab:retrieval_priors}.

\section{Possible stellar heterogeneities} \label{app:stellar}

The star TOI-270 is a quiet M dwarf, showing no significant signs of rotational modulation, spot crossings, or flare events in TESS photometry, along with displaying low H$\alpha$ activity \citep{Gunther2019}. Nevertheless, we considered the effects of unocculted stellar heterogeneities on the observed transmission spectrum using our AURA retrieval framework \citep{Pinhas2018}. We did not find evidence for stellar heterogeneities in our canonical one-offset case, given that a model without these effects is preferred at 2.3$\sigma$. This is consistent with similar analysis with previous HST WFC3 observations \cite{Evans2023}. To investigate the extent to which this is driven by the WFC3 data, obtained at a different epoch, we performed another retrieval with stellar heterogeneities included using only the present NIRSpec G395H data. From the NIRSpec G395H observation alone, we again found no evidence for stellar heterogeneities and a model without stellar heterogeneities is preferred at 2.2$\sigma$, supporting the claims that TOI-270 is a particularly quiet M dwarf. In this case, we find that the heterogeneity covering fraction $f_\mathrm{het}$ and temperature $T_\textrm{het}$ are constrained to $f_\mathrm{het} = 0.08^{+0.05}_{-0.08}$ and $T_\textrm{het} = 3357^{+190}_{-163}$, respectively. 
Including the WFC3 data does not significantly change these stellar heterogeneity constraints. Finally, we note that the chemical abundance constraints are not significantly affected by the consideration of stellar heterogeneities.

\section{Transmission spectrum of TOI 270~b} \label{app:TOI_270b}

\begin{figure}
        \includegraphics[width=0.493\textwidth]{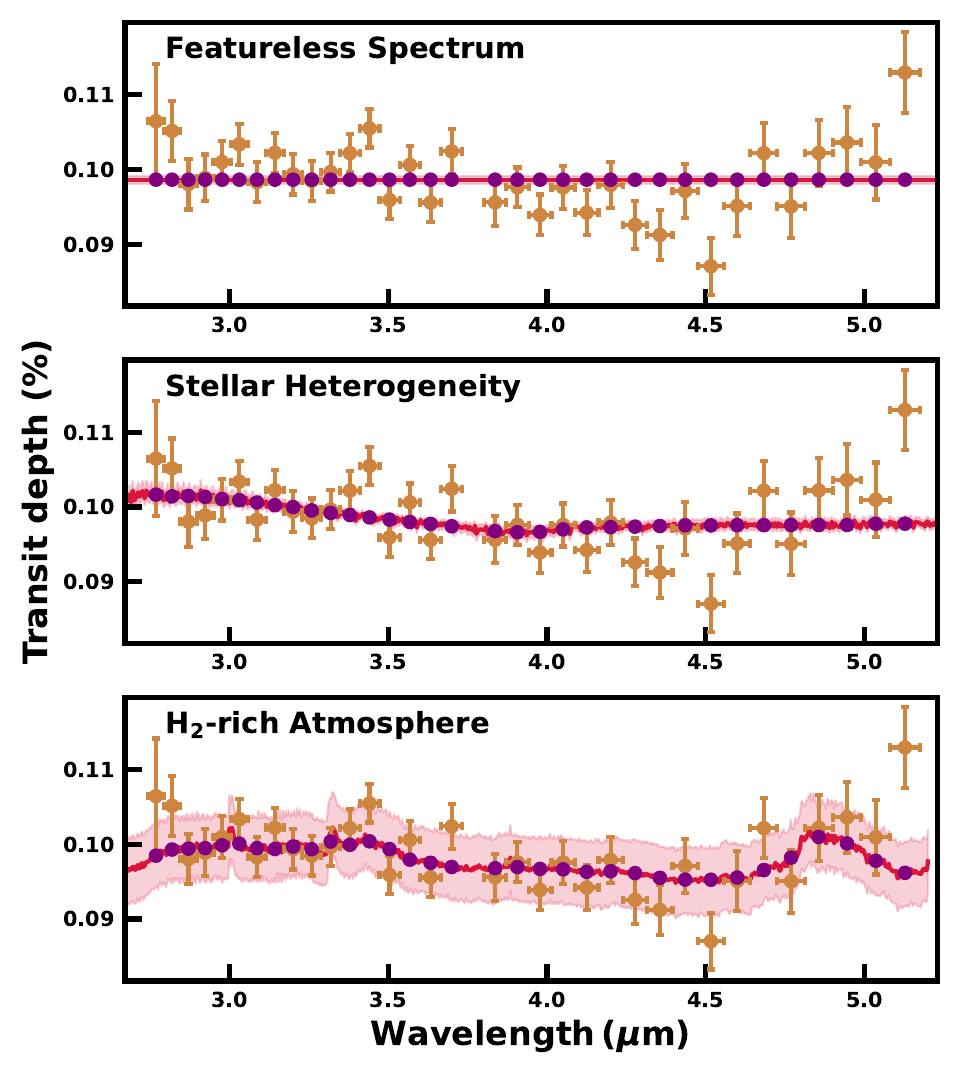}
  \vspace{-5mm}
    \caption{Transmission spectrum of TOI-270~b observed with JWST/NIRSpec G395H. The spectrum is binned to $R \approx 55$ for visual clarity and is shown in orange (same in all panels). The three panels show three model scenarios used to fit the observed spectrum. For the stellar heterogeneity case, we show the fit using the uninformative priors. The observed spectrum of TOI-270~b shows tentative spectral features; however, more observations are needed to confirm these. The red curves show the median retrieved model spectra, while the lighter-red contours denote the 1$\sigma$ intervals. The purple points correspond to the median spectrum binned to match the observations.
    }
    \label{fig:T270b} 
\end{figure}

We present the transmission spectrum of the super-Earth TOI-270~b. With a size of 1.2-1.3 $R_\oplus$ and orbiting close-in around an M-dwarf, it is not known whether the planet could retain a H$_2$-rich atmosphere on Gyr time scales. The observed spectrum shows some spectral features that are consistent with a H$_2$-rich atmosphere; however, at the present noise level, there is a risk of over-fitting the data. More observations are required to confidently determine the nature of the planet's atmosphere and to rule out effects due to stellar heterogeneities and other systematics. However, the prospects of an H$_2$-rich atmosphere on TOI-270~b are certainly interesting, as it would open the door for detailed atmospheric characterisation of rocky exoplanets with JWST.

To explain the observed spectrum, we considered three scenarios: (1) a featureless planetary spectrum, (2) a featureless planetary spectrum contaminated by stellar heterogeneities, and (3) a spectrum with molecular absorption originating from a H$_2$-rich atmosphere. For the latter case, we adopted the same retrieval framework as for TOI-270~d (described in Section \ref{sec:retrieval}). The NIRSpec G395H transmission spectrum of TOI-270~b is shown in Figure \ref{fig:T270b}, together with the model fits from each of the three scenarios. For the stellar heterogeneity case, we consider two sets of priors on the heterogeneity coverage fraction, $f_\mathrm{het}$, and heterogeneity temperature, $T_{\textrm{het}}$, either with uninformative priors ($f_\mathrm{het}$ uniform between 0-0.5, and $T_{\textrm{het}}$ uniform between 2300-4207 K) or informative priors ($f_\mathrm{het}$ normally distributed with a mean of 0.08 and a standard deviation of 0.08, truncated at 0 and 0.5, and $T_{\textrm{het}}$ normally distributed with a mean of 3357 K and a standard deviation of 190 K) as constrained by the NIRSpec spectrum of TOI-270~d (see Appendix \ref{app:stellar}). We used a stellar effective temperature of $3506\pm70$ K \citep{VanEylen2021}. In the first two scenarios, we used uniform priors on the transit depth, using a $\pm200$ ppm range around the value from \cite{VanEylen2021}.

We found that the spectrum of TOI-270~b deviates from a featureless spectrum at a significance of 3.3-3.5$\sigma$ when compared to either a stellar heterogeneity model (uninformative priors) or a H$_2$-rich atmosphere model. This demonstrates the degeneracy between these two scenarios, as recently reported by \cite{moran_high_2023} in the case of GJ 486~b. However, when using informative priors on the heterogeneity coverage fraction and temperature, we found that the preference of the stellar heterogeneity model over a featureless spectrum is lowered to 2.6$\sigma$. This corresponds to a 2.7$\sigma$ preference for the H$_2$-rich atmosphere scenario compared to the informed stellar heterogeneity case. We conclude that there is marginal evidence for a H$_2$-rich atmosphere on TOI-270~b. We note that within this context, the spectral feature at $\sim$5$ \mu$m could potentially be explained by OCS \citep{Madhusudhan2021}. Ultimately, more observations are needed to robustly constrain the presence and composition of the planet's atmosphere and to rule out systematics.

\section{Comparison with K2-18~b} \label{app:K2_18b}

\begin{figure*}[h]
        \includegraphics[width=\textwidth]{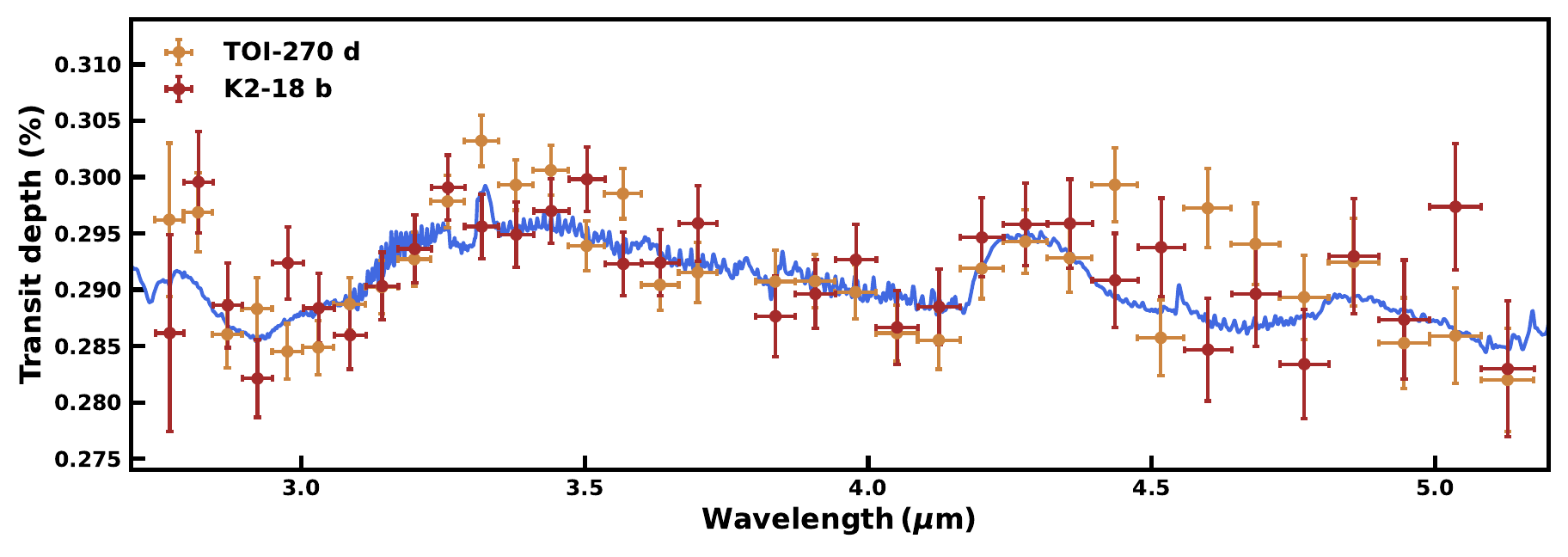}
  \vspace{-5mm}
    \caption{Comparison between the transmission spectrum of TOI-270~d and K2-18~b observed with JWST/NIRSpec G395H. The spectra are binned to $R \approx 55$ for visual clarity and are shown in orange and dark red, respectively. The K2-18~b spectrum is vertically offset by -41 ppm as discussed in \cite{Madhu_carbon2023}. The two spectra show similar spectral features, coming predominantly from the absorption due to CH$_4$ and CO$_2$. The blue curves shows the median retrieved model spectrum obtained with the K2-18~b data from \cite{Madhu_carbon2023}.
    }
    \label{fig:K2_18b_comparison} 
\end{figure*}

We compared the current NIRSpec G395H spectrum of the Hycean candidate TOI-270~d to the recent spectrum of K2-18~b, observed with the same instrument \citep{Madhu_carbon2023}. We show the two spectra overlaid in Figure \ref{fig:K2_18b_comparison} together with the K2-18~b model fit. Both spectra show remarkably similar spectral features, with the most prominent absorption coming from the same carbon-bearing species, namely CH$_4$ and CO$_2$. For both planets, the abundance constraints of CH$_4$ and CO$_2$ nominally show median abundances of $\sim$0.1-1\% with uncertainties smaller than 1 dex. However, we note that the CO$_2$ abundance constraint for TOI-270~d varies with the treatment of limb darkening, as discussed in Appendix \ref{app:LDCs}. As in the case of K2-18~b, we do not find significant contributions from NH$_3$ in the atmosphere of TOI-270~d, indicating that the atmosphere of TOI-270~d is seemingly inconsistent with a deep-atmosphere, mini-Neptune scenario. 

The retrieved temperature of TOI-270~d, namely,  $T_{\textrm{10mbar}} = 305^{+101}_{-93}$ K in the canonical one-offset case, is somewhat higher than that of K2-18~b, with $T_{\textrm{10mbar}} = 242^{+79}_{-57}$ K (for the one-offset case in \cite{Madhu_carbon2023}). This is consistent with the tentative evidence for H$_2$O in the atmosphere of TOI-270~d, given that the inferred temperature range is high enough to allow for H$_2$O vapour at the low pressures probed in transmission spectroscopy. At the same time, we did not constrain clouds or hazes in TOI-270~d in the observable atmosphere. This is in contrast to K2-18~b, where clouds and hazes are somewhat preferred by the data and where no evidence of H$_2$O was found. These findings suggest a seemingly consistent picture in which H$_2$O may be condensing into clouds on K2-18~b and not on TOI-270~d due to the temperature difference. However, although the $T_{\textrm{10mbar}}$ constraint for K2-18 b allows for a cold-trap, with temperatures below 200 K at 1$\sigma$, it also allows for significantly warmer temperatures comparable to TOI-270~d. Therefore, further observations are needed to more robustly constrain the presence of H$_2$O and clouds/hazes, and their temperature dependence. Upcoming NIRISS observations of TOI-270~d, in the 0.6-2.8 $\mu$m range, will be particularly valuable in this pursuit (JWST GO Programs 2759 and 4098). 

\section{Atmospheric retrieval details} \label{app:retrieval}

\begin{table*}
\small
\def\arraystretch{1.4}
\caption{Retrieved atmospheric parameters and prior probability distributions.}
\vspace{-2mm}
\begin{tabular}{lclccccc} \hline \hline
    Parameter & Bayesian Prior  & Description & One offset & Two offsets & CLD & DT & NIRSpec only\\[0.5mm]
    \hline
    $\mathrm{log}(X_\mathrm{H_2O})$& $\mathcal{U}$(-12, -0.3) & Mixing ratio of H$_2$O & $-3.92_{-4.43}^{+1.70}$ & $-5.98_{-3.75}^{+2.79}$ & $-1.04_{-0.45}^{+0.24}$ & $-1.91_{-0.94}^{+0.57}$ & $-4.80_{-4.21}^{+2.09}$\\[0.5mm]
    $\mathrm{log}(X_\mathrm{CH_4})$  & $\mathcal{U}$(-12, -0.3) & Mixing ratio of CH$_4$ & $-2.44_{-0.46}^{+0.34}$ & $-2.70_{-0.49}^{+0.42}$ & $-2.97_{-0.39}^{+0.30}$ & $-2.72_{-0.50}^{+0.41}$ & $-2.42_{-0.53}^{+0.40}$\\[0.5mm]
    $\mathrm{log}(X_\mathrm{NH_3})$  & $\mathcal{U}$(-12, -0.3) & Mixing ratio of NH$_3$ & $-8.86_{-1.94}^{+2.00}$ & $-9.16_{-1.80}^{+1.90}$ & $-8.58_{-2.10}^{+2.20}$ & $-8.95_{-1.88}^{+1.98}$ & $-8.91_{-1.91}^{+1.99}$\\[0.5mm]
    $\mathrm{log}(X_\mathrm{CO})$ & $\mathcal{U}$(-12, -0.3) & Mixing ratio of CO & $-4.98_{-4.03}^{+2.52}$ & $-4.14_{-4.23}^{+1.93}$ & $-7.20_{-2.94}^{+2.66}$ & $-7.20_{-2.96}^{+3.03}$ & $-4.96_{-4.20}^{+2.49}$\\[0.5mm]
    $\mathrm{log}(X_\mathrm{CO_2})$& $\mathcal{U}$(-12, -0.3) & Mixing ratio of CO$_2$ & $-1.96_{-0.79}^{+0.49}$ & $-2.01_{-0.73}^{+0.50}$ & $-3.95_{-0.90}^{+0.72}$ & $-2.46_{-0.92}^{+0.71}$ & $-1.93_{-0.77}^{+0.49}$ \\[0.5mm]
    $\mathrm{log}(X_\mathrm{HCN})$  & $\mathcal{U}$(-12, -0.3)& Mixing ratio of HCN & $-8.84_{-1.94}^{+2.13}$ & $-9.28_{-1.71}^{+1.94}$ & $-8.23_{-2.36}^{+2.38}$ & $-8.86_{-1.97}^{+2.20}$ & $-8.96_{-1.89}^{+2.12}$ \\[0.5mm]
    $\mathrm{log}(X_\mathrm{C_2H_2})$  & $\mathcal{U}$(-12, -0.3) & Mixing ratio of C$_2$H$_2$ & $-9.31_{-1.65}^{+1.73}$ & $-9.55_{-1.53}^{+1.59}$ & $-8.85_{-1.96}^{+2.02}$ & $-9.31_{-1.64}^{+1.80}$ & $-9.38_{-1.63}^{+1.73}$ \\[0.5mm]
    $\mathrm{log}(X_\mathrm{C_2H_6})$  & $\mathcal{U}$(-12, -0.3) & Mixing ratio of C$_2$H$_6$ & $-2.65_{-1.41}^{+0.97}$ & $-3.48_{-2.08}^{+1.23}$ & $-3.31_{-2.09}^{+1.12}$ & $-4.16_{-4.34}^{+1.70}$ & $-2.53_{-1.45}^{+0.98}$ \\[0.5mm]
    $\mathrm{log}(X_\mathrm{H_2S})$  & $\mathcal{U}$(-12, -0.3) & Mixing ratio of H$_2$S & $-6.16_{-3.57}^{+3.25}$  & $-5.61_{-3.93}^{+2.93}$ & $-5.74_{-3.78}^{+2.93}$ & $-5.12_{-4.12}^{+2.51}$ & $-6.65_{-3.39}^{+3.45}$ \\[0.5mm]
    $\mathrm{log}(X_\mathrm{SO_2})$  & $\mathcal{U}$(-12, -0.3) & Mixing ratio of SO$_2$ & $-8.03_{-2.35}^{+2.31}$ & $-7.55_{-2.75}^{+2.26}$ & $-7.92_{-2.47}^{+2.13}$ & $-8.00_{-2.45}^{+2.18}$ & $-8.14_{-2.38}^{+2.30}$ \\[0.5mm]
    $\mathrm{log}(X_\mathrm{DMS})$  & $\mathcal{U}$(-12, -0.3) & Mixing ratio of DMS & $-8.69_{-2.02}^{+2.04}$ & $-8.83_{-1.98}^{+1.97}$ & $-7.01_{-3.04}^{+1.53}$ & $-6.76_{-3.16}^{+1.26}$ & $-8.75_{-2.04}^{+2.00}$ \\[0.5mm]
    $\mathrm{log}(X_\mathrm{CS_2})$& $\mathcal{U}$(-12, -0.3) & Mixing ratio of CS$_2$ & $-2.59_{-0.95}^{+0.67}$ & $-2.63_{-0.94}^{+0.66}$ & $-4.13_{-1.31}^{+0.92}$ & $-3.07_{-0.91}^{+0.74}$ & $-2.67_{-1.06}^{+0.73}$ \\[0.5mm]
    $\mathrm{log}(X_\mathrm{CH_3Cl})$  & $\mathcal{U}$(-12, -0.3)& Mixing ratio of CH$_3$Cl & $-7.83_{-2.55}^{+2.62}$ & $-7.36_{-2.83}^{+2.58}$ & $-7.33_{-2.82}^{+2.52}$ & $-8.04_{-2.46}^{+2.46}$ & $-8.02_{-2.44}^{+2.56}$ \\[0.5mm]
    $\mathrm{log}(X_\mathrm{OCS})$ & $\mathcal{U}$(-12, -0.3)& Mixing ratio of OCS & $-8.42_{-2.15}^{+2.06}$ & $-8.28_{-2.29}^{+2.07}$ & $-8.89_{-1.93}^{+1.93}$ & $-6.97_{-2.49}^{+1.39}$ & $-8.41_{-2.22}^{+2.07}$ \\[0.5mm]
    $\mathrm{log}(X_\mathrm{N_2O})$ & $\mathcal{U}$(-12, -0.3) & Mixing ratio of N$_2$O & $-8.80_{-1.94}^{+2.07}$ & $-9.00_{-1.88}^{+2.05}$ & $-9.03_{-1.84}^{+1.96}$ & $-8.94_{-1.88}^{+1.95}$ & $-8.87_{-1.95}^{+2.13}$ \\[0.5mm]
    $T_0 / \mathrm{K} $ & $\mathcal{U}$(0, 500)& Temperature at 1 $\mu$bar  & $219_{-83}^{+101}$ & $188_{-68}^{+87}$ & $308_{-132}^{+99}$ & $199_{-71}^{+76}$ & $186_{-83}^{+107}$ \\[0.5mm]
    $T_{10 \mathrm{mbar}} / \mathrm{K}$ & - & Temperature at 10 mbar & $305_{-93}^{+101}$ & $268_{-73}^{+87}$ & $374_{-107}^{+104}$  & $289_{-75}^{+80}$ & $282_{-90}^{+109}$ \\[0.5mm]
    $\alpha_1 / \mathrm{K}^{-\frac{1}{2}}$  & $\mathcal{U}$(0.02, 2.00) & $P$-$T$ profile curvature & $1.19_{-0.44}^{+0.47}$ & $1.20_{-0.45}^{+0.48}$ & $1.28_{-0.49}^{+0.45}$ & $1.17_{-0.46}^{+0.49}$ & $1.19_{-0.45}^{+0.49}$ \\[0.5mm]
    $\alpha_2/ \mathrm{K}^{-\frac{1}{2}}$& $\mathcal{U}$(0.02, 2.00)  & $P$-$T$ profile curvature & $1.04_{-0.53}^{+0.57}$ & $1.04_{-0.52}^{+0.58}$ & $1.05_{-0.56}^{+0.58}$ & $1.01_{-0.56}^{+0.59}$ & $0.97_{-0.54}^{+0.61}$ \\[0.5mm]

    $\mathrm{log}(P_1/\mathrm{bar})$   & $\mathcal{U}$(-6, 1) & $P$-$T$ profile region limit & $-1.97_{-1.43}^{+1.27}$ & $-2.00_{-1.45}^{+1.31}$ & $-1.98_{-1.45}^{+1.29}$ & $-1.99_{-1.44}^{+1.28}$ & $-2.03_{-1.43}^{+1.30}$ \\[0.5mm]
    $\mathrm{log}(P_2/\mathrm{bar})$  & $\mathcal{U}$(-6, 1) & $P$-$T$ profile region limit & $-4.21_{-1.17}^{+1.50}$ & $-4.23_{-1.17}^{+1.56}$ & $-4.19_{-1.19}^{+1.52}$ & $-4.24_{-1.15}^{+1.56}$ & $-4.29_{-1.11}^{+1.51}$ \\[0.5mm]
    $\mathrm{log}(P_3/\mathrm{bar})$   & $\mathcal{U}$(-2, 1)& $P$-$T$ profile region limit & $-0.11_{-0.94}^{+0.72}$ & $-0.11_{-0.94}^{+0.73}$ & $-0.12_{-0.93}^{+0.73}$ & $-0.11_{-0.96}^{+0.74}$ & $-0.14_{-0.92}^{+0.75}$ \\[0.5mm]
    $\mathrm{log}(P_\mathrm{ref}/\mathrm{bar})$   & $\mathcal{U}$(-6, 1) & Reference pressure at R$_\mathrm{P}$ & $-5.32_{-0.39}^{+0.46}$ & $-5.24_{-0.45}^{+0.51}$ & $-4.67_{-0.50}^{+0.48}$ & $-4.97_{-0.54}^{+0.58}$ & $-5.51_{-0.34}^{+0.35}$ \\[0.5mm]
    $\mathrm{log}(a)$  & $\mathcal{U}$(-4, 10)& Rayleigh enhancement factor & $1.71_{-3.59}^{+4.15}$ & $1.71_{-3.58}^{+4.42}$ & $1.52_{-3.51}^{+4.19}$ & $1.61_{-3.54}^{+4.28}$ & $2.17_{-3.91}^{+4.45}$ \\[0.5mm]
    $\gamma$   & $\mathcal{U}$(-20, 2)& Scattering slope & $-10.27_{-6.12}^{+7.08}$ & $-10.28_{-6.25}^{+7.36}$ & $-10.69_{-6.01}^{+7.00}$ & $-10.43_{-6.12}^{+7.24}$ & $-10.50_{-5.90}^{+6.54}$ \\[0.5mm]
    $\mathrm{log}(P_\mathrm{c}/\mathrm{bar})$  & $\mathcal{U}$(-6, 1)& Cloud top pressure & $-1.56_{-2.49}^{+1.58}$ & $-1.54_{-2.77}^{+1.62}$ & $-1.37_{-2.13}^{+1.47}$ & $-1.78_{-2.18}^{+1.72}$ & $-1.21_{-2.40}^{+1.37}$  \\[0.5mm]
    $\phi$  & $\mathcal{U}$(0, 1)& Cloud/haze coverage fraction & $0.31_{-0.20}^{+0.30}$ & $0.27_{-0.18}^{+0.33}$ & $0.29_{-0.19}^{+0.35}$ & $0.31_{-0.20}^{+0.30}$ & $0.33_{-0.22}^{+0.35}$ \\[0.5mm]
    $\delta_\mathrm{NIRSpec} / \mathrm{ppm}$  & $\mathcal{U}$(-200, 200)& NIRSpec dataset offset & $-84_{-14}^{+15}$  & - & $-84_{-14}^{+14}$ & $-75_{-14}^{+14}$ & -\\[0.5mm]
    $\delta_\mathrm{NRS1} / \mathrm{ppm}$  & $\mathcal{U}$(-200, 200)& NIRSpec NRS1 dataset offset & - & $-91_{-15}^{+15}$ & - & - & -\\[0.5mm]
    $\delta_\mathrm{NRS2} / \mathrm{ppm}$ & $\mathcal{U}$(-200, 200) & NIRSpec NRS2 dataset offset & - & $-73_{-16}^{+16}$ & - & - & -\\[0.5mm]
    \hline
\end{tabular}
\vspace{2mm}
\newline
\footnotesize{\textbf{Note.} The five retrieval cases are described in Section \ref{sec:retrieval}. All parameter values correspond to the retrieved median and 1$\sigma$ estimates. The temperature at 10~mbar, which corresponds to the observed photosphere, is derived from the retrieved $P-T$ profile. Additional parameters for stellar heterogeneities are described in Appendix \ref{app:retrieval}.
}
\label{tab:retrieval_priors}
\end{table*}

Our atmospheric model, described in Section \ref{sec:retrieval}, encompasses 27 independent parameters in the canonical retrieval case. These parameters, shown in Table \ref{tab:retrieval_priors}, are as follows: 15 correspond to the individual mixing ratios of the included chemical species, 6 are assigned to the representation of the $P$-$T$ profile, 4 pertain to the modelling of clouds and hazes, and 1 to the reference pressure denoted as $P_\mathrm{ref}$. This reference pressure is defined as the pressure occurring at a fixed planetary radius of 2.241 $R_\oplus$ for TOI-270~d. These parameters are described in more detail in \cite{Pinhas2018}. Furthermore, depending on the retrieval case, we also considered an offset on either the full NIRSpec spectrum (relative to WFC3) or on NRS1 and NRS2 individually, adding either one or two more parameters, similar to \cite{Madhu_carbon2023}. The retrieval was conducted on the native resolution NIRSpec G395H spectrum ($R \sim 2700$).

\begin{figure*}[h]
        \includegraphics[width=\textwidth]{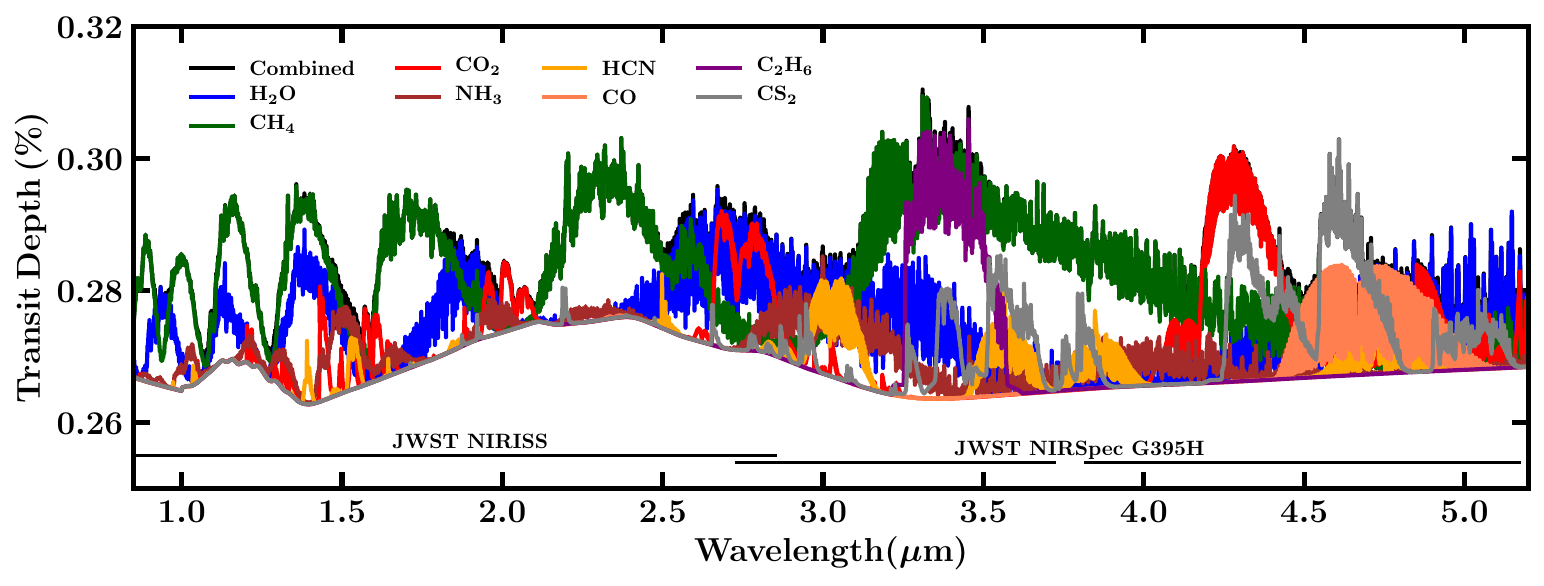}
  \vspace{-5mm}
    \caption{Spectral contributions of key molecular species in the 1-5 $\mu$m range. The different curves show individual contributions from different molecules to a nominal model transmission spectrum of TOI-270~d shown in black and denoted as Combined. The model assumes a mixing ratio of 10$^{-2}$ for CH$_4$, CO$_2$, and H$_2$O, 10$^{-3}$ for C$_2$H$_6$ and CS$_2$, and 10$^{-5}$ for all the other species, consistent with our retrieval estimates, and an isothermal temperature profile of 300 K. Each curve corresponds to a transmission spectrum with opacity contributions from a single molecule at a time, in addition to H$_2$-H$_2$ and H$_2$-He collision-induced absorption. The spectral ranges of NIRISS SOSS and NIRSpec G395H are indicated.
    }
    \label{fig:contribution} 
\end{figure*}

In terms of the chemical species, the retrievals include prominent CNOS species expected in temperate H$_2$-rich atmospheres: H$_2$O, CH$_4$, NH$_3$, HCN, CO, CO$_2$, C$_2$H$_2$, C$_2$H$_6$, H$_2$S, and SO$_2$. The molecular absorption cross sections were obtained from the HITRAN and Exomol databases: H$_2$O \citep{Polyansky2018}, CH$_4$ \citep{Hargreaves2020}, NH$_3$ \citep{Coles2019}, HCN \citep{harris2006, barber2014}, CO \citep{rothman2010, Li2015}, CO$_2$ \citep{HUANG2013, huang2017}, C$_2$H$_2$ \citep{Chubb2020}, C$_2$H$_6$ \citep{Gordon2020}, H$_2$S \citep{Azzam2016, Chubb2018}, and SO$_2$ \citep{Underwood2016}. 
The absorption cross sections are computed following the methods of \cite{Gandhi2017} and \cite{gandhi2020}, including thermal broadening and pressure broadening due to H$_2$ or air as available. We also considered several additional molecules that have been suggested to be promising biomarkers in habitable rocky exoplanets \citep{Segura2005, domagal-goldman2011, seager2013a, seager2013b, catling2018, schwieterman2018} as well as Hycean worlds \citep{Madhusudhan2021}: (CH$_3$)$_2$S (or DMS), CS$_2$, CH$_3$Cl, OCS, and N$_2$O. The absorption cross-sections of CH$_3$Cl, OCS, CS$_2$, and N$_2$O are  computed from the corresponding line lists from the \textsc{HITRAN} database \citep{HITRAN2020}: CH$_3$Cl \citep{ch3cl_1, ch3cl_2}, OCS \citep{ocs_1, ocs_2, ocs_3, ocs_4, ocs_5, ocs_6, ocs_7}, CS$_2$ \citep{CS2-gamma_air-1-1299, CS2-nu-2-1298, CS2-nu-3-1297, CS2-nu-4-1296}, and N$_2$O \citep{n2o_2}. For DMS, we used the absorption cross sections provided directly by HITRAN \citep{Sharpe2024, Gordon2017, Kochanov2019}. We note that the molecular line lists for these less prominent species are currently limited and more experimental and theoretical work in the future could improve on the line data available, especially for H$_2$-rich atmospheres. The mean molecular weight (MMW) of the atmosphere is derived from the volume mixing ratios of the molecules for a given model evaluation. In Figure \ref{fig:contribution}, we show the spectral contributions of key molecular species in a representative model transmission spectrum of TOI-270~d.  We also considered the clouds, hazes and stellar heterogeneity as described in \cite{Pinhas2018}.

In Table \ref{tab:retrieval_priors}, we list all atmospheric model parameters and the prior probability distributions used in the five retrieval cases (described in Section \ref{sec:retrieval}). For the stellar heterogeneity parameters, we use the priors $\mathcal{N}$(3506 K, 100$^2$ K$^2$) (K), $\mathcal{U}$(1753 K, 4207 K), and $\mathcal{U}$(0, 0.5), for the stellar photosphere temperature, stellar heterogeneity temperature, and coverage fraction, respectively. Here $\mathcal{U}(X, Y)$ denotes a uniform probability distribution between X and Y, while $\mathcal{N}(\mu, \sigma^2)$ denotes a normal distribution of mean $\mu$ and variance $\sigma^2$.

\end{appendix}

\end{document}